\documentstyle[12pt]{article}

\begin{document}

\newcommand{\G}{{\cal G}}

\begin{titlepage}

\begin{flushright}
{BI-TP-94/13}\\
{UQMATH-94-02}
{hep-th/9405030}
\end{flushright}
\vspace{10mm}
\begin{center}
{\LARGE On the construction of trigonometric solutions}\\
{\LARGE of the Yang-Baxter equation}\\
\vspace{11mm}
{\large Gustav W. Delius}\\
\vspace{5mm}
Fakult\"at f\"ur Physik, Universit\"at Bielefeld\\
Postfach 10 01 31, D-33501 Bielefeld, Germany\\
{\small e-mail: delius@physf.uni-bielefeld.de}\\
\vspace{8mm}
{\large Mark D. Gould } and {\large Yao-Zhong Zhang}\\
\vspace{5mm}
Department of Mathematics, University of Queensland\\
Brisbane Qld 4072, Australia.\\
{\small e-mail: yzz@maths.uq.oz.au}\\
\vspace{6mm}
April 1994

\end{center}

\begin{abstract}
We describe the construction of trigonometric R-matrices corresponding to
the (multiplicity-free) tensor product of two irreducible representations
of a quantum algebra $U_q(\G)$. Our method is a generalization of the tensor
product graph method to the case of two different representations.
It yields the decomposition of the R-matrix
into projection operators. Many new examples of trigonometric R-matrices
(solutions to the spectral parameter dependent Yang-Baxter equation) are
constructed using this approach.
\end{abstract}

\end{titlepage}

\renewcommand{\baselinestretch}{1.3} 


\newcommand{\sect}[1]{\setcounter{equation}{0}\section{#1}}
\renewcommand{\theequation}{\thesection.\arabic{equation}}


\newcommand{\reff}[1]{eq.~(\ref{#1})}


\newcommand{\beq}{\begin{equation}}
\newcommand{\eeq}{\end{equation}}
\newcommand{\bea}{\begin{eqnarray}}
\newcommand{\eea}{\end{eqnarray}}
\newcommand{\bq}{\begin{quote}}
\newcommand{\eq}{\end{quote}}

\newtheorem{Theorem}{Theorem}
\newtheorem{Definition}{Definition}
\newtheorem{Lemma}[Theorem]{Lemma}
\newtheorem{Corollary}[Theorem]{Corollary}
\newcommand{\proof}[1]{{\bf Proof. }
        #1\begin{flushright}$\Box$\end{flushright}}


\parskip 0.3cm


\newcommand{\half}{\frac{1}{2}}
\renewcommand{\P}{{\cal P}}
\newcommand{\e}{{\epsilon}}
\newcommand{\C}{\bf C}
\renewcommand{\a}{\alpha}
\renewcommand{\b}{\beta}
\renewcommand{\l}{\lambda}
\newcommand{\s}{\sigma}
\newcommand{\D}{\Delta}
\newcommand{\uqg}{$U_q({\G})$}
\newcommand{\uqgh}{$U_q(\hat{\G})$}
\newcommand{\qh}{q^{h_0/2}}
\newcommand{\qmh}{q^{-h_0/2}}
\newcommand{\cq}{$U_q(C)$}
\newcommand{\cqh}{$U_q(\hat{C})$}
\newcommand{\gq}{$U_q(G)$}
\newcommand{\gqh}{$U_q(\hat{G})$}

\sect{Introduction}

The solutions to the Yang-Baxter equation play a central role in the
theory of quantum integrable models \cite{Skl79,Fad81,Kor94}.
In statistical mechanics they are the
Boltzman weights of exactly solvable lattice models \cite{Bax82}. In
quantum field theory they give the exact factorizable scattering
matrices \cite{Zam79}. For an introduction to the mathematical aspects
of the Yang-Baxter equation see e.g. \cite{Jim89}.

The Yang-Baxter equation (with additive spectral parameter) has the form
\beq
R_{12}(\theta)R_{13}(\theta+\theta')R_{23}(\theta')
  =R_{23}(\theta')R_{13}(\theta'+\theta)R_{12}(\theta).
\label{YB}
\eeq
The $R_{ab}(\theta)$ are matrices which depend on a spectral parameter
$\theta$ and which act on the tensor product
of two vector spaces $V_a$ and $V_b$
\beq
R_{ab}(\theta):~V_a\otimes V_b\rightarrow V_a\otimes V_b.
\eeq
The products of $R$'s in \reff{YB} act on the space $V_1\otimes V_2
\otimes V_3$.

In this paper we are interested in solutions of \reff{YB} which are
trigonometric functions of the spectral parameter $\theta$ and which
depend on a further complex parameter $q$. These solutions arise
as the intertwiners of the quantum affine algebras $U_q(\hat{\G})$
\cite{Dri85,Jim85,Dri86,Jim86a,Fre92}. These are deformations of affine
Kac-Moody algebras
\cite{Kac90}. Associated to any two finite-dimensional irreducible
$U_q(\hat{\G})$-modules\footnote{For a construction of the
finite-dimensional irreducible representations of $U_q(\hat{\G})$ see
\cite{Del94}.}
$V(\l)$ and $V(\mu)$ there exists a trigonometric
R-matrix $R^{\l\mu}(\theta)$. Given three modules, the R-matrices for all
pairs of these three modules are a solution of \reff{YB}.

We will describe a method for constructing such R-matrices.
The method works for any untwisted quantum affine algebra $U_q(\hat{\G})$
and for all those irreducible finite dimensional $U_q(\hat{\G})$-modules
$V(\l)$ and $V(\mu)$ which
are irreducible also as modules of the associated non-affine quantum algebra
$U_q(\G)$ and whose tensor product $V(\l)\otimes V(\mu)$ is a multiplicity-free
direct sum of irreducible $U_q(\G)$ modules.
The method is a generalization of \cite{Zha91} which treats the special
case $V(\l)=V(\mu)$.

Let us summarize the previous work on trigonometric R-matrices. Most is
known if $V(\l)=V(\mu)$ coincides with the defining representation. The
corresponding R-matrices have been constructed for $A_n^{(1)}$
\cite{Che80}, $A_2^{(2)}$ \cite{Ize81}, $A_2^{(1)}$ \cite{Zam80},
and finally for all members of the classical series $A_n^{(1)},~
B_n^{(1)},~C_n^{(1)},~D_n^{(1)},~A_{2n}^{(2)},~A_{2n-1}^{(2)}$
and $D_n^{(2)}$ \cite{Baz85,Jim86b}. $E_6^{(1)}$ and $E_7^{(1)}$ were
treated in \cite{Ma90}. The R-matrices in higher representations
can be obtained by applying the fusion procedure
\cite{Kul81} to the trigonometric R-matrices.
In principle this method yields the R-matrices for the product of
any two representations which
can be projected out of the multiple tensor product of fundamental
representations, using the Hecke algebra in the case of $A_n$
\cite{Jim86a,Cha91a} or the Birman-Wenzl-Murakami algebra in the case of
$B_n, C_n,$ and $D_n$ \cite{Mac92b}. Because
of the algebraic complications this has been done in practice only
for the product of the vector with a two-index tensor representation
\cite{Mac92b}.

More practical is the method of \cite{Zha91} which is designed for the
multiplicity-free tensor product $V(\l)\otimes V(\l)$ of some irrep with
itself.
In \cite{Zha91} the R-matrices in such a special case have been worked out for
the symmetric as well as the antisymmetric tensor representations of $A_n$,
the spinor represenations for $D_n$, the 27-dimensional representation
of $E_6$ and the 54-dimensional representation of $E_7$.

An alternative general method for the construction of
trigonometric R-matrices is to start with the formula for
the universal R-matrix of $U_q(\hat{\G})$ \cite{Tol92,Kho92,Zha93a}
and to specialize it to particular representations \cite{Zha93b,Bra94}.
This method becomes impractical for larger representations, although it
has obvious advantages when multiplicities occur in the tensor product
space.

More is known
\cite{Ber78,Kul81,Ogi86,Ogi87,Res87,Cha90,Cha91b,Mac91a,Mac91b,Mac92a}
about solutions of the Yang-Baxter equation
(\ref{YB}) which are rational functions of the spectral parameter
$\theta$ and which do not depend on an extra parameter $q$.
They arise as the interwiners of tensor products of
finite dimensional representations of the Yangians \cite{Dri85,Dri86}.
These rational R-matrices can be obtained from the trigonometric
R-matrices, which are the subject of this paper, by taking the limit
$q\rightarrow 1$.

The paper is organized as follows: In section \ref{jimboeqs} we
review  quantum affine algebras and Jimbo equations which arise
from them and which uniquely determine trigonometric solutions of the
Yang-Baxter equation (\ref{YB}). In section \ref{general} we show how
to write the general solution to these equations as a sum over
twisted projection operators and we give the equations which uniquely
determine the coefficients. In many cases it is possible to solve
these equations using only classical Lie algebra representation theory
and we do this using  tensor product graphs in section \ref{TPG}.
There we explicitly treat various examples and determine the following
R-matrices:
\begin{description}
\item[$A_n$:] i) rank $a$ symmetric tensor with
rank $b$ symmetric tensor; ii) rank $a$ symmetric tensor with
rank $b$ antisymmetric tensor; and iii) rank $a$
antisymmetric tensor with rank $b$ antisymmetric tensor.
\item[$B_n$:] i) spinor times spinor, ii) symmetric traceless tensor
of rank $a$ times symmetric traceless tensor of rank $b$.
\item[$C_n$:] fundamental representation $a$ with fundamental
representation $b$.
\item[$D_n$:] i) spinor times antispinor, ii) symmetric traceless tensor
of rank $a$ times symmetric traceless tensor of rank $b$.
\item[$E_6$:] i) $V(\l_1)$ times $V(a\l_1)$, ii) $V(\l_5)$ times
$V(a\l_1)$.
\item[$F_4$:] $V(\l_4)$ times $V(\l_4)$.
\end{description}
We end with a discussion in section \ref{discussion}.

\sect{Quantum Affine Algebras and the Jimbo Equations\label{jimboeqs}}
Let $A=(a_{ij})_{0\leq i,j\leq r}$ be a symmetrizable
generalized Cartan matrix in the sense of Kac \cite{Kac90} and
let $\hat{\cal G}$ denote the affine Lie algebra
associated with the Cartan  matrix $A$,
where $a_{ij}=2(\alpha_i,
\alpha_j)/(\alpha_i,\alpha_i)$,
$\alpha_i$ are the simple roots of $\hat{\cal G}$ and
$r$ is the rank of the corresponding finite-dimensional simple Lie algebra
${\cal G}$. We are restricting our attention to the untwisted affine
Lie algebras.
The quantum affine algebra $U_q(\hat{\cal G})$ \cite{Dri85,Jim85,Dri86} is
defined by
generators $\{e_i,~f_i,~q^{h_i},~i=0,1,...,r\}$ and relations
\begin{eqnarray}
&&q^h.q^{h'}=q^{h+h'}\,,~~~~h,~ h'=h_i,~ i=0,1,...,r\nonumber\\
&&q^he_iq^{-h}=q^{(h,\alpha_i)} e_i\,,~~q^hf_iq^{-h}=q^{-(h,\alpha_i)}
  f_i\nonumber\\
&&[e_i, f_j]=\delta_{ij}\frac{q^{h_i}-q^{-h_i}}{q-q^{-1}}\nonumber\\
&&\sum^{1-a_{ij}}_{k=0}(-1)^k e_i^{(1-a_{ij}-k)}e_je_i^{(k)}
   =0~~~~(i\neq j)\nonumber\\
&&\sum^{1-a_{ij}}_{k=0}(-1)^k f_i^{(1-a_{ij}-k)}f_jf_i^{(k)}
   =0~~~~(i\neq j)\label{relations1}
\end{eqnarray}
where
\begin{equation}
e_i^{(k)}=\frac{e^k_i}{[k]_{q_i}!},~~~f^{(k)}_i=\frac{f^k_i}{[k]_{q_i}!}
\,,~~~[k]_q=\frac{q^k-q^{-k}}{q-q^{-1}},~~~q_i=q^{(\alpha_i,\alpha_i)/2},~~~
[k]_q!=\prod_{n=1}^k [n]_q.
\end{equation}
Here we do not include the derivation $d$ as part of $U_q(\hat{\cal G})$.
Furthermore we will set the central charge to zero.

The associated non-affine quantum algebra $U_q(\G)$ is generated by
$\{e_i,~f_i,~q^{h_i},~i=1,...,r\}$ (i.e., without $e_0$, $f_0$ and $h_0$)
and the same relations.
The algebra $U_q(\hat{\cal G})$ is a Hopf algebra with coproduct, counit and
antipode similar to the case of $U_q(\cal G)$:
\begin{eqnarray}
&&\Delta(q^h)=q^h\otimes q^h\,,~~~h=h_i,~~~i=0,1,\cdots, r\nonumber\\
&&\Delta(e_i)=e_i\otimes q^{h_i/2}+q^{-h_i/2}\otimes e_i\nonumber\\
&&\Delta(f_i)=f_i\otimes q^{h_i/2}+q^{-h_i/2}\otimes f_i\nonumber\\
&&S(a)=-q^{h_\rho}aq^{-h_\rho}\,,~~\forall a\in U_q({\G})\label{coproduct}
\end{eqnarray}
where $\rho$ is the half-sum of positive roots of $\G$.
We have omitted the remaining formulas for the antipode and the
counit since they are not required.

Let $\D'$ be the opposite coproduct: $\D'=T\D$, where $T$ is
the twist map: $T(a\otimes b)=b\otimes a\,,~\forall a,b\in U_q(\hat{\G})$.
Then $\Delta$ and $\Delta'$ are related by the universal R-matrix $R$
in $U_q(\hat{\G})\otimes U_q(\hat{\G})$ satisfying, among others,
the relations
\begin{eqnarray}
&&R\D(a)=\D'(a)R\,,~~~~~\forall a\in U_q(\hat{\G})\label{intw}\\
&&(I\otimes \D)R=R_{13}R_{12}\,,~~~~(\D\otimes I)R=R_{13}R_{23}\label{hopf}
\end{eqnarray}
where if $R=\sum a_i\otimes b_i$ then $R_{12}=\sum a_i\otimes b_i\otimes 1$
, $R_{13}=\sum a_i\otimes 1\otimes b_i$ etc.

For any $x\in{\C}$, we define an automorphism $D_x$ of \uqgh{} as \cite{Fre92}
\beq
D_x(e_i)=x^{\delta_{i0}}e_i\,,~~~~D_x(f_i)=x^{-\delta_{i0}}f_i,
{}~~~~D_x(h_i)=h_i
\eeq
and set
\beq
R(x)=(D_x\otimes I)(R).
\eeq
It follows from (\ref{hopf}) that $R(x)$ solves the QYBE with multiplicative
spectral parameter
\beq
R_{12}(x)R_{13}(xy)R_{23}(y)=R_{23}(y)R_{13}(xy)R_{12}(x).\label{qybe}
\eeq
This reproduces \reff{YB} if we set $x=e^\theta$, $y=e^{\theta'}$.

Let $\pi_\l\,,~\pi_\mu$ and $\pi_\nu$ be three arbitrary
finite-dimensional irreps of
\uqg{} afforded by the irreducible modules $V(\l)\,,~V(\mu)$ and $V(\nu)$
respectively, where $\l\,,~\mu$ and $\nu$ are the highest
weights of these modules. Assume all $\pi_\l\,,~\pi_\mu$ and $\pi_\nu$ are
affinizable, i.e. they can be extended to finite dimensional irreps of the
corresponding \uqgh~\footnote{For an investigation of this point see
\cite{Del94}.}. Let
\beq
R^{\l\mu}(x)=g_q(x)(\pi_\l\otimes \pi_\mu)(R(x))
\eeq
where $g_q(x)$ is a scalar normalization factor (see appendix
\ref{u-condition}). Then $R^{\l\mu}(x)$ satisfies the system of linear
equations \cite{Jim86b} deduced from the intertwining property (\ref{intw})
\bea
&&R^{\l\mu}(x)\D^{\l\mu}(a)=\D^{'\l\mu}(a)R^{\l\mu}(x)~~~~~
  \forall a\in U_q({\cal G}),\nonumber\\
&&R^{\l\mu}(x)\left (x\pi_\l(e_0)\otimes \pi_\mu(\qh)+\pi_\l(\qmh)\otimes
  \pi_\mu(e_0)\right )\nonumber\\
&&~~~~~  =\left (x\pi_\l(e_0)\otimes \pi_\mu(\qmh)+\pi_\l(\qh)\otimes
  \pi_\mu(e_0)\right )R^{\l\mu}(x),\label{jimboeq1}
\eea
where $\D^{\l\mu}(a)\equiv (\pi_\l\otimes\pi_\mu)\D(a)$. The solution
$R^{\l\mu}(x)$ to the above equations intertwines the coproduct
of $U_q(\hat{\G})$ acting on the
$U_q(\hat{\G})$-module $V(\l)\otimes V(\mu)$. Because the representation
$(\pi_\l\otimes\pi_\mu)(D_x\otimes 1)$ is irreducible for generic $x$, by
Schur's lemma the solution is uniquely determined up to the scalar factor
$g_q(x)$. $R^{\l\mu}(x)$ satisfies the QYBE in
$V(\nu)\otimes V(\l)\otimes V(\mu)$
\beq
R^{\nu\l}(x)R^{\nu\mu}(xy)R^{\l\mu}(y)=R^{\l\mu}(y)
  R^{\nu\mu}(xy)R^{\nu\l}(x).\label{qybe1}
\eeq

Now let $P^{\l\mu}: V(\l)\otimes V(\mu)\rightarrow V(\mu)\otimes V(\l)$
be the permutation operator such that
\beq
P^{\l\mu}(|f>\otimes |g>)=|g>\otimes |f>,~~~~~~\forall |f>\in V(\l),~ |g>\in
  V(\mu)
\eeq
and denote
\beq
{\bf\check{R}}^{\l\mu}(x)=P^{\l\mu}\,R^{\l\mu}(x)\,,
\eeq
then (\ref{qybe1}) is translated into the following relations:
\beq
({\bf\check{R}}^{\l\mu}(x)\otimes I)(I\otimes {\bf\check{R}}^{\nu\mu}(xy))
  ({\bf\check{R}}^{\nu\l}(y)\otimes I)=(I\otimes {\bf\check{R}}^{\nu\l}(y))
  ({\bf\check{R}}^{\nu\mu}
  (xy)\otimes I)(I\otimes {\bf\check{R}}^{\l\mu}(x))\label{qybe2}
\eeq
and eqs.(\ref{jimboeq1}) can be rewritten as
\bea
&&{\bf\check{R}}^{\l\mu}(x)\D^{\l\mu}(a)=\D^{\mu\l}(a)
  {\bf\check{R}}^{\l\mu}(x)~~~~~
  \forall a\in U_q({\cal G}),\label{jimboeq2a}\\
&&{\bf\check{R}}^{\l\mu}(x)\left (x\pi_\l(e_0)\otimes \pi_\mu(\qh)+\pi_\l(\qmh)
  \otimes\pi_\mu(e_0)\right )\nonumber\\
&&~~~~~  =\left (\pi_\mu(e_0)\otimes \pi_\l(\qh)+x\pi_\mu(\qmh)\otimes
  \pi_\l(e_0)\right ){\bf\check{R}}^{\l\mu}(x),\label{jimboeq2}
\eea
Eqs.(\ref{jimboeq2a}, \ref{jimboeq2})
are the so-called Jimbo equations (JEs). In this paper
we present a method for solving these equations.

\sect{General solution to Jimbo equations\label{general}}

Below we assume that $V(\l)\otimes V(\mu)$ is a multiplicity-free direct sum
of irreducible modules. We choose to normalize the solution so that
\beq
{\bf\check{R}}^{\l\mu}(x){\bf\check{R}}^{\mu\l}(x^{-1})=I.\label{unitary2}
\eeq
See appendix \ref{u-condition} for an account of this.

We first focus on the $x=1$ case of the JEs which take the form
\bea
\label{inter}
&&{\bf\check{R}}^{\l\mu}(1)\D^{\l\mu}(a)=
  \D^{\mu\l}(a){\bf\check{R}}^{\l\mu}(1)~~~~~
  \forall a\in U_q({\cal G}),\\
&&{\bf\check{R}}^{\l\mu}(1)\left (\pi_\l(e_0)\otimes
  \pi_\mu(\qh)+\pi_\l(\qmh)\otimes
  \pi_\mu(e_0)\right )\nonumber\\
&&~~~~~  =\left (\pi_\mu(e_0)\otimes \pi_\l(\qh)+\pi_\mu(\qmh)\otimes
  \pi_\l(e_0)\right ){\bf\check{R}}^{\l\mu}(1).\label{r-1}
\eea
In this case we have
\beq
(R^T)^{\l\mu}(1)R^{\l\mu}(1)=I=
{\bf\check{R}}^{\l\mu}(1){\bf\check{R}}^{\mu\l}(1).
\eeq

We decompose the tensor product $V(\l)\otimes V(\mu)$
into its irreducible
$U_q({\G})$-submo\-dules. Due to the similarity between the representation
theory of the quantum algebra and the classical algebra $\G$, this
decomposition is identical to the classical one \cite{Ros88,Lus88}
\beq
V(\l)\otimes V(\mu)=\bigoplus_\nu  V(\nu)\label{free}
\eeq
where the sum is over the highest weights occuring in the tensor product space
(assumed multiplicity-free).

Let $\P^{\l\mu}_\nu:~V(\l)\otimes V(\mu)\rightarrow V(\nu)$ be the
projection operators which satisfy
\beq
\P^{\l\mu}_\nu\;\P^{\l\mu}_{\nu'}=\delta_{\nu\nu'}
  \P^{\l\mu}_\nu\,,~~~~\sum_{\nu}\P^{\l\mu}_\nu=I.
\eeq
We now define operators ${\bf\check{P}}^{\l\mu}_\nu$ by
\beq
{\bf\check{P}}^{\l\mu}_\nu
=\P^{\mu\l}_\nu\;{\bf\check{R}}^{\l\mu}(1)={\bf\check{R}}^{\l\mu}(1)
  \;\P^{\l\mu}_\nu.
\eeq
Then, by definition
\beq
{\bf\check{R}}^{\l\mu}(1)=\sum_{\nu}{\bf\check{P}}^{\l\mu}_\nu.
  \label{solution-1}
\eeq
It is easy to show that
\beq
{\bf\check{P}}^{\l\mu}_\nu\;\P^{\l\mu}_{\nu'}=\P^{\mu\l}_{\nu'}\;
  {\bf\check{P}}^{\l\mu}_\nu
  =\delta_{\nu\nu'}{\bf\check{P}}^{\l\mu}_\nu,\label{projection1}
\eeq
and
\bea
{\bf\check{P}}^{\mu\l}_\nu\;{\bf\check{P}}^{\l\mu}_{\nu'}&=&\P^{\l\mu}_\nu
{\bf\check{R}}^{\mu\l}(1){\bf\check{R}}^{\l\mu}(1)\P^{\l\mu}_{\nu'}\nonumber\\
  &=&\P^{\l\mu}_\nu\P^{\l\mu}_{\nu'}=\delta_{\nu\nu'}
  \P^{\l\mu}_\nu\label{projection2}
\eea
Eqs.(\ref{projection1},~\ref{projection2}) imply that the operators
${\bf\check{P}}^{\l\mu}_\nu$ are ``projection" operators.
As can be seen from \reff{inter}, the "projectors"
${\bf\check{P}}^{\l\mu}_\nu$ are intertwiners of $U_q(\G)$.
The general solution of \reff{jimboeq2a} is a sum of these elementary
intertwiners
\beq
{\bf\check{R}}^{\l\mu}(x)=\sum_{\nu}\rho_\nu(x)\;
  {\bf\check{P}}^{\l\mu}_\nu\label{r-general}
\eeq
where $\rho_\nu(x)$ are some functions of $x$.
According to \reff{solution-1} they satisfy
\beq\label{initial}
\rho_\nu(1)=1.
\eeq
Our task is now to determine these functions so that \reff{r-general}
satisfies also Jimbo's equation \reff{jimboeq2}.

We begin by determining $\rho_\nu(x)$ at the special value $x=0$.
At $x=0$ the R-matrix ${\bf\check{R}}^{\l\mu}(0)$
reduces to the R-matrix of the quantum algebra
$U_q({\G})$ in $V(\l)\otimes V(\mu)$:
${\bf\check{R}}^{\l\mu}(0)\equiv {\bf\check{R}}^{\l\mu}$.
We recall the following fact about the universal R-matrix $\bar{R}$ for
\uqg (see also appendix \ref{u-condition}):
\beq
\bar{R}^T\bar{R}=(v\otimes v)\D(v^{-1})\,,~~~v=uq^{-2h_\rho}\,,~~~
  \pi_\l(v)=q^{-C(\l)}\cdot I_{V(\l)}
\eeq
with $u=\sum_iS(\bar{b}_i)\bar{a}_i$ where
$\bar{R}=\sum_i\bar{a}_i\otimes\bar{b}_i$. $S$ is the antipode for \uqg,
$C(\l)=(\l,\l+2\rho)$
is the eigenvalue of the quadratic Casimir element of $\G$ in an irrep with
highest weight $\l$;~ $\rho$ is the half-sum of positive roots of $\G$.
We now compute ${\bf\check{R}}^{\mu\l}{\bf\check{R}}^{\l\mu}$. Using the above
equations we have
\bea
\sum_\nu(\rho_\nu(0))^2\P^{\l\mu}_\nu&=&
  {\bf\check{R}}^{\mu\l}{\bf\check{R}}^{\l\mu}\equiv
  P^{\mu\l}\bar{R}^{\mu\l}P^{\l\mu}\bar{R}^{\l\mu}\nonumber\\
&=&(\bar{R}^T)^{\l\mu}\bar{R}^{\l\mu}=(\pi_\l\otimes\pi_\mu)(\bar{R}^T
  \bar{R})\nonumber\\
&=&\pi_\l(v)\otimes\pi_\mu(v)(\pi_\l\otimes\pi_\mu)\D(v^{-1})\nonumber\\
&=&\sum_{\nu} q^{C(\nu)-C(\l)-C(\mu)}\P^{\l\mu}_\nu
\eea
where use has been made of
${\bf\check{R}}^{\l\mu}\equiv P^{\l\mu}\bar{R}^{\l\mu}
\equiv P^{\l\mu}(\pi_\l\otimes \pi_\mu)(\bar{R})$. It follows immediately that
\beq\label{rho0}
\rho_\nu(0)=\e(\nu)q^{\frac{C(\nu)-C(\l)-C(\mu)}{2}}
\eeq
where $\e(\nu)$ is the parity of $V(\nu)$ in $V(\l)\otimes V(\mu)$ (see
appendix \ref{parity} for an explanation). Thus
\beq
{\bf\check{R}}^{\l\mu}(0)=\sum_{\nu}\e(\nu)
q^{\frac{ C(\nu)-C(\l)-C(\mu)}{2}}
  {\bf\check{P}}^{\l\mu}_\nu.
\eeq
A similar relation was obtained in \cite{Res88} using a different approach.
\vskip.1in
\noindent{\bf Remark:} (i) although irrelevant to this paper, it is worth
pointing out that in the case of a tensor product decomposition with finite
multiplicities a similar spectral decomposition formula for ${\bf\check{R}}
^{\l\mu}(0)$ may be obtained by combining the techniques above and those in
\cite{Gou92,Gou93}; (ii) given the universal $R$-matrix $\bar{R}$
for $U_q(\G)$ (which
is known in explicit form for any $\G$ \cite{KR91}) then ${\bf\check{R}
}^{\l\mu}(0)={\bf\check{R}}^{\l\mu}$ can be independently computed by
using ${\bf\check{R}}^{\l\mu}(0)=P^{\l\mu}(\pi_\l\otimes\pi_\mu)(\bar{R})$.
Once ${\bf\check{R}}^{\l\mu}(0)$ is known through such a independent way,
then from the above equation we have
\beq
{\bf\check{P}}^{\l\mu}_\nu=\e(\nu)
q^{-\frac{ C(\nu)-C(\l)-C(\mu)}{2}}
{\bf\check{R}}^{\l\mu}(0)\P^{\l\mu}_\nu
\eeq
which implies that ${\bf\check{P}}^{\l\mu}_\nu$ (and therefore ${\bf\check{R}}^
{\l\mu}(1)$) are computable this way. ${\bf\check{P}}^{\l\mu}_\nu$ can also
be determined by using pure representation
theory of $U_q(\G)$. This is explained in an forthcoming paper \cite{DGZ94}
(for a brief outline of this see appendix \ref{projectors}). ~~~~$\Box$

Now we use Jimbo's equation \reff{jimboeq2} to determine the
$\rho_\nu(x)$ for general $x$. We insert
(\ref{r-general}) into (\ref{jimboeq2}),
multiply the resulting equation by $\P^{\mu\l}_\nu$ from the left and
by $\P^{\l\mu}_{\nu'}$ from the right and use the properties of the
projectors. We arrive at
\bea
&&\rho_\nu(x){\bf\check{P}}^{\l\mu}_\nu\left (x\pi_\l(e_0)\otimes
  \pi_\mu(\qh)+\pi_\l(\qmh)
  \otimes\pi_\mu(e_0)\right )\P^{\l\mu}_{\nu'}\nonumber\\
&&~~~~~  =\rho_{\nu'}(x)\P^{\mu\l}_\nu\left (\pi_\mu(e_0)\otimes
  \pi_\l(\qh)+x\pi_\mu(\qmh)\otimes
  \pi_\l(e_0)\right ){\bf\check{P}}^{\l\mu}_{\nu'}.\label{2}
\eea
To simplify this we use the following equation, obtained by taking $x=0$
and using \reff{rho0}:
\bea
&&\e(\nu)q^{C(\nu)/2}{\bf\check{P}}^{\l\mu}_\nu\left
(\pi_\l(\qmh)\otimes\pi_\mu
  (e_0)\right )\P^{\l\mu}_{\nu'}\nonumber\\
&&~~~~~ =\e(\nu')q^{C(\nu')/2}\P^{\mu\l}_\nu\left (\pi_\mu(e_0)\otimes
  \pi_\l(\qh)
  \right ){\bf\check{P}}^{\l\mu}_{\nu'}.\label{p-0}
\eea
We get a second useful relation by multiplying
the above by ${\bf\check{P}}^{\mu\l}_\nu$ from the left and
by ${\bf\check{P}}^{\mu\l}_{\nu'}$ from the right and then interchanging the
labels $\mu$ and $\l$ to give
\bea
&&\e(\nu)q^{-C(\nu)/2}{\bf\check{P}}^{\l\mu}_\nu\left (
  \pi_\l(e_0)\otimes\pi_\mu(\qh)
  \right )\P^{\l\mu}_{\nu'}\nonumber\\
&&~~~~~  =\e(\nu')
  q^{-C(\nu')/2}\P^{\mu\l}_\nu\left (\pi_\mu(\qmh)\otimes \pi_\l(e_0)\right )
 {\bf\check{P}}^{\l\mu}_{\nu'}.
  \label{p-infty}
\eea
A third important relation follows if we set $x=1$ in \reff{2} and use
\reff{initial}:
\bea
&&{\bf\check{P}}^{\l\mu}_\nu\left (\pi_\l(e_0)\otimes
  \pi_\mu(\qh)+\pi_\l(\qmh)
  \otimes\pi_\mu(e_0)\right )\P^{\l\mu}_{\nu'}\nonumber\\
&&~~~~~  =\P^{\mu\l}_\nu\left (\pi_\mu(e_0)\otimes
  \pi_\l(\qh)+\pi_\mu(\qmh)\otimes
  \pi_\l(e_0)\right ){\bf\check{P}}^{\l\mu}_{\nu'}.\label{p-1}
\eea
In view of eq.(\ref{p-0}), eq.(\ref{p-infty}) and eq.({\ref{p-1})
we obtain from (\ref{2}), for $\nu\neq\nu'$,
\bea
&&\rho_\nu(x)\left (xq^{C(\nu)/2}+\e(\nu)\e(\nu')q^{C(\nu')/2}\right )
  {\bf\check{P}}^{\l\mu}_\nu\left (\pi_\l(e_0)\otimes \pi_\mu(\qh)\right )
  \P^{\l\mu}_{\nu'}\nonumber\\
&&~~ =\rho_{\nu'}(x)\left (q^{C(\nu)/2}+\e(\nu)\e(\nu')xq^{C(\nu')/2}
  \right ){\bf\check{P}}^{\l\mu}_\nu\left (\pi_\l(e_0)\otimes \pi_\mu(\qh)
  \right )\P^{\l\mu}_{\nu'}\,.
\eea
Multiplying by ${\bf\check{P}}^{\mu\l}_\nu$ from the left we see that, if
\beq\label{cond1}
\P^{\l\mu}_\nu\left (\pi_\l(e_0)\otimes \pi_\mu(\qh)\right )
  \P^{\l\mu}_{\nu'}\neq 0\label{me}\,,~~~~\nu\neq\nu'
\eeq
the above equation gives
rise to a relation between $\rho_\nu(x)$ and $\rho_{\nu'}(x)$
\beq
\rho_\nu(x)=\rho_{\nu'}(x)\,\frac{q^{C(\nu)/2}
  +\e(\nu)\e(\nu')xq^{C(\nu')/2}}
  {xq^{C(\nu)/2}+\e(\nu)\e(\nu')q^{C(\nu')/2}}\,,~~~~\nu\neq\nu'\,.
\label{general-solution}
\eeq
Note that this relation is unchanged under $\nu\leftrightarrow \nu'$, as it
must be because also the condition \reff{cond1} has this invariance.
It simplifies if we note that $\epsilon(\nu)\epsilon(\nu')=-1$ whenever
the condition \reff{cond1} is satisfied. This is explained in appendix
\ref{parity}.
With the notation
\beq
\langle a\rangle\equiv\frac{1-x q^{a}}{x-q^{a}},
\eeq
the relation \reff{general-solution} then becomes
\beq\label{rel}
\rho_\nu(x)=\left\langle \frac{C(\nu')-C(\nu)}{2}\right\rangle\,
\rho_{\nu'}(x)\,,~~~~\nu\neq\nu'.
\eeq

\sect{Tensor Product Graphs and R-matrices\label{TPG}}

We have a relation \reff{rel} between the coefficients
$\rho_\nu$ and $\rho_{\nu'}$ whenever the condition
\reff{cond1} is satisfied, i.e., whenever $\pi_\l(e_0)\otimes \pi_\mu(\qh)$
maps from the module $V(\nu')$ to the module $V(\nu)$.
As a graphical aid \cite{Zha91} we introduce the tensor product graph.

\begin{Definition}\label{reducibilitygraph}
The {\bf tensor product graph} $G^{\l\mu}$ associated to the tensor product
$V(\l)\otimes V(\mu)$ is a graph
whose vertices are the irreducible modules
$V(\nu)$ appearing in the decomposition \reff{free} of
$V(\l)\otimes V(\mu)$. There is an edge between a vertex $V(\nu)$
and a vertex $V(\nu')$ iff
\beq\label{link}
\P^{\l\mu}_\nu\left (\pi_\l(e_0)\otimes \pi_\mu(\qh)\right )
  \P^{\l\mu}_{\nu'}\neq 0.
\eeq
\end{Definition}
For an example of a tensor product graph see figure \ref{bspin}.

If $V(\l)$ and $V(\mu)$ are irreducible $U_q(\G)$-modules the
tensor product graph is always connected, i.e., every node is linked to every
other node by a path of edges. This follows from the fact that
$\pi_\l(e_0)\otimes \pi_\mu(\qh)$ is related to the lowest component
of an adjoint tensor operator; for details see \cite{Zha91}. This implies
that the relations \reff{general-solution} are sufficient to
determine all the coefficients $\rho_\nu(x)$ uniquely.
If the tensor product graph is multiply connected, i.e., if there exists
more than two paths between two nodes, then the relations overdetermine
the coefficients, i.e., there are consistency conditions \cite{Zha91}.
However, because the existence of a solution to JEs is guaranteed by
the existence of the universal R-matrix, these consistency conditions
will always be satisfied.

The straightforward but tedious and impractical way to determine the tensor
product graph is to work out explicitly the left hand side  of (\ref{link}).
To do so requires a knowledge of the
representations $\pi_\l(e_0)$ and $\pi_\mu(e_0)$ as well as the projection
operators $\P^{\l\mu}_\nu$. Although it may be possible to construct these
representations \cite{Del94} and the projectors explicitly, it is more
practical to use other approaches.

One is to consider the case of $q=1$ and to work with the following smaller
graph.

\begin{Definition}\label{subreduciblitygraph}
The {\bf restricted tensor product graph} $G^{(0)\l\mu}$ associated to the
tensor product $V^{(0)}(\l)\otimes V^{(0)}(\mu)$, where $V^{(0)}(\l)$ etc.
denote irreducible $\G$-modules, is a graph
whose vertices are the irreducible modules
$V^{(0)}(\nu)$ appearing in the decomposition \reff{free} of
$V^{(0)}(\l)\otimes V^{(0)}(\mu)$. There is an edge between a vertex
$V^{(0)}(\nu)$ and a vertex $V^{(0)}(\nu')$ iff
\beq\label{sublink}
\P^{(0)\l\mu}_\nu\left (\pi^{(0)}_\l(e_0)\otimes I_\mu)\right )
  \P^{(0)\l\mu}_{\nu'}\neq 0.
\eeq
where $\P^{(0)\l\mu}_\nu$  are the $q=1$ versions of the projectors
$\P^{\l\mu}_\nu$.
\end{Definition}
Following similar arguments to \cite{Zha91} it is seen that this restricted
tensor product graph is in general "smaller" than the tensor product graph,
i.e. it may be missing some edges. It is however still connected and thus
imposing a relation \reff{general-solution} for each of the edges of the
restricted tensor product graph will lead to a unique solution.
It is thus sufficient to determine the restricted tensor product graph.
Unfortunately this too is a difficult task.

Another approach is to work instead with the following larger graph which we
will utilize in this paper.

\begin{Definition}\label{tpg}
The {\bf extended tensor product graph} $\Gamma^{\l\mu}$ associated to the
tensor product $V(\l)\otimes V(\mu)$ is a graph
whose vertices are the irreducible modules
$V(\nu)$ appearing in the decomposition \reff{free} of
$V(\l)\otimes V(\mu)$. There is an edge between two vertices $V(\nu)$
and $V(\nu')$ iff
\beq
V(\nu')\subset V_{adj}\otimes V(\nu) ~~~\mbox{ and }~~
\epsilon(\nu)\epsilon(\nu')=-1.\label{adj}
\eeq
\end{Definition}
It follows again from the fact that $\pi_\l(e_0)\otimes \pi_\mu(\qh)$
is related to the lowest component of an adjoint tensor operator that
the condition \reff{adj} is a necessary condition for \reff{link}
\cite{Zha91}.
This means that every link contained in the tensor product graph is contained
also in the extended tensor product graph but the latter may contain
more links. Only if the extended tensor product graph is a tree do we
know that it is equal to the tensor product graph. If we impose a
relation (\ref{rel}) on the $\rho$'s
for every link in the extended tensor product graph, we may be
imposing too many relations and thus may not always find a solution.
If however we do
find a solution, then this is the unique correct solution which we
would have obtained also from the tensor product graph.

The advantage of using the extended tensor product graph is that it can be
constructed using only Lie algebra representation theory.
We only need to be able to decompose tensor products and to determine
the parity of submodules. The decomposition of tensor products can
be done for small representations by matching dimensions and Casimir
values. In the general case one can use Young tableau techniques
or characters.

In the following we will give
a number of examples of tensor product graphs. We are interested only in
the tensor product graphs for products of those
irreducible $U_q(\G)$-modules which are affinizable, i.e., which are
also modules of the quantum
affine algebra $U_q(\hat{\G})$. All so-called
miniscule modules are affinizable. These are modules for which all
$SU(2)$ submodules are either trivial or two-dimensional. Also those
modules are affinizable which correspond to extremal nodes on a tensor product
graph of two other affinizable modules \cite{Del94}.

\subsection{Examples for $A_n=sl(n+1)$}

We introduce the standard orthonormal weight basis $\{\e_i|i=1,\cdots,n\}$ for
$A_n$. The fundamental weights are
\beq
\l_b=\sum_{i=1}^b\e_i\,,~~~~~b=1,2,\cdots,n
\eeq
and the semisum of positive roots is
\beq
\rho=\sum_{i=1}^n(n-2i+1)\e_i\,.
\eeq
All representations of $U_q(A_n)$ are affinizable and therefore there is
an unlimited number
of tensor product graphs which give rise to R-matrices. Here
we only consider three simple cases: (i) rank $a~(\geq 1)$ symmetric tensor
(which has the highest weight $a\l_1$) with
rank $b~(\geq 1)$ symmetric tensor; (ii) rank $a$ symmetric tensor with
rank $b~(1\leq b\leq n)$ antisymmetric tensor (which has the fundamental weight
$\l_b$ as its highest weight); and (iii) rank $a$
antisymmetric tensor
with rank $b$ antisymmetric tensor ($1\leq a,b\leq n$). Without
loss of generality, we assume $a\geq b$ in the following. We start with
case (i). Its tensor product graph is
\beq
\unitlength=1mm
\linethickness{0.4pt}
\begin{picture}(112.60,7.60)(10,12)
\put(50.00,15.00){\circle*{5.20}}
\put(65.00,15.00){\circle*{5.20}}
\put(95.00,15.00){\circle*{5.20}}
\put(110.00,15.00){\circle*{5.20}}
\put(44.00,15.00){\makebox(0,0)[rc]{$V(a\l_1)\otimes V(b\l_1)~=$}}
\put(50.00,11.00){\makebox(0,0)[ct]{$\Lambda_0$}}
\put(65.00,11.00){\makebox(0,0)[ct]{$\Lambda_1$}}
\put(95.00,11.00){\makebox(0,0)[ct]{$\Lambda_{b-1}$}}
\put(110.00,11.00){\makebox(0,0)[ct]{$\Lambda_b$}}
\put(110.00,15.00){\line(-1,0){21.00}}
\put(50.00,15.00){\line(1,0){21.00}}
\put(80.00,15.00){\makebox(0,0)[cc]{$\cdots$}}
\end{picture}
\eeq
where
\beq
\Lambda_c=(a+b-c)\e_1+c\e_2=(a+b-2c)\l_1+c\l_2,~~~~~c=0,1,\cdots,b.
\eeq
The Casimir takes the following values on the representations appearing in the
above graph
\beq
C(\Lambda_c)=(a+b-c)^2+c^2-2c+(a+b)(n-1).
\eeq
The R-matrix is
\beq
{\bf\check{R}}^{a\l_1,b\l_1}(x)=\rho_{\Lambda_0}(x)
\sum_{c=0}^b\prod_{i=1}^c\left\langle
  a+b-2i+2\right\rangle\,{\bf\check{P}}^{a\l_1,b\l_1}_{\Lambda_c}
\eeq
(our convention here and below is that $\prod_{i=1}^0(\dots)=1$.).
The case of $a=b$ was worked out in \cite{Zha91}.
The overall scalar factor $\rho_{\Lambda_0}(x)$ is not
determined by the Jimbo equations or by the Yang-Baxter equation but
by the normalization condition \reff{unitary2}. It satisfies
\reff{initial} and \reff{rho0}. We will from now on drop such overall
scalar factors from the formulas for the R-matrices.

For case (ii) we have the tensor product graph
\beq
\unitlength=1mm
\linethickness{0.4pt}
\begin{picture}(77.60,7.60)(10,12)
\put(50.00,15.00){\circle*{5.20}}
\put(44.00,15.00){\makebox(0,0)[rc]{$V(a\l_1)\otimes V(\l_b)~=$}}
\put(50.00,11.00){\makebox(0,0)[ct]{$\Lambda_1$}}
\put(75.00,15.00){\circle*{5.20}}
\put(50.00,15.00){\line(1,0){25.00}}
\put(75.00,11.00){\makebox(0,0)[ct]{$\Lambda_2$}}
\end{picture}
\eeq
where
\beq
\Lambda_1=(a+1)\e_1+\sum_{i=2}^b\e_i=a\l_1+\l_b,~~~~~
  \Lambda_2=a\e_1+\sum_{i=2}^{b+1}\e_i=(a-1)\l_1+\l_{b+1}.
\eeq
The difference of the Casimirs is
\beq
\frac{C(\Lambda_1)-C(\Lambda_2)}{2}=a+b
\eeq
The R-matrix is
\beq
{\bf\check{R}}^{a\l_1,\l_b}(x)={\bf\check{P}}^{a\l_1,\l_b}_{\Lambda_1}
  +\left\langle a+b\right\rangle\,{\bf\check{P}}^{a\l_1,\l_b}_{\Lambda_2}
\eeq

Finally for case (iii) the tensor product graph is
\beq
\unitlength=1mm
\linethickness{0.4pt}
\begin{picture}(112.60,7.60)(10,12)
\put(50.00,15.00){\circle*{5.20}}
\put(65.00,15.00){\circle*{5.20}}
\put(95.00,15.00){\circle*{5.20}}
\put(110.00,15.00){\circle*{5.20}}
\put(44.00,15.00){\makebox(0,0)[rc]{$V(\l_a)\otimes V(\l_b)~=$}}
\put(50.00,11.00){\makebox(0,0)[ct]{$\Lambda_0$}}
\put(65.00,11.00){\makebox(0,0)[ct]{$\Lambda_1$}}
\put(95.00,11.00){\makebox(0,0)[ct]{$\Lambda_{b-1}$}}
\put(110.00,11.00){\makebox(0,0)[ct]{$\Lambda_{min(b, n-a)}$}}
\put(110.00,15.00){\line(-1,0){21.00}}
\put(50.00,15.00){\line(1,0){21.00}}
\put(80.00,15.00){\makebox(0,0)[cc]{$\cdots$}}
\end{picture}
\eeq
where
\beq
\Lambda_c=\sum_{i=1}^c\e_i+\sum_{i=1}^{a+b-c}\e_i\equiv\sum_{i=1}^{a+c}\e_i
  +\sum_{i=1}^{b-c}\e_i=\l_{a+c}+\l_{b-c},~~~~~c=0,1,\cdots,min(b, n-a)
\eeq
The Casimirs are
\beq
C(\Lambda_c)=2c(b-a-1)-2c^2+(n+1)(a+b)-a^2-b^2
\eeq
and the R-matrix is
\beq
{\bf\check{R}}^{\l_a,\l_b}(x)=\sum_{c=0}^{min(b, n-a)}\prod_{i=1}^c\left\langle
 2i+a-b\right\rangle\,{\bf\check{P}}
 ^{\l_a,\l_b}_{\Lambda_c}\,.
\eeq
Again the case of $a=b$ was worked out in \cite{Zha91}. This R-matrix has
been used by Hollowood \cite{Hol93a} to construct the soliton S-matrix in
$A_n^{(1)}$ Toda theory.

\subsection{Examples for $B_n=so(2n+1)$}

In the orthonormal weight basis $\{\e_i|i=1,\dots,n\}$ for the weight space
of $B_n$ the fundamental weights are
\beq
\l_n=\half\sum_{i=1}^n \e_i~,~~~~
\l_a=\sum_{i=1}^a \e_i~,~~a=1,\dots,n-1.
\eeq
The semisum of positive roots is
\beq
\rho=\sum_{i=1}^n\,\left(n-i+\half\right)\,\e_i.
\eeq

We begin with the spinor representation $V(\l_n)$ because it is the
only miniscule representation of $B_n$. Its tensor product graph is
shown in figure \ref{bspin}.

\begin{figure}[ht]
\unitlength=1.00mm
\linethickness{0.4pt}
\begin{picture}(110.00,48.00)
\put(70.00,40.00){\circle*{5.20}}
\put(80.00,30.00){\circle*{5.20}}
\put(95.00,15.00){\circle*{5.20}}
\put(105.00,5.00){\circle*{5.20}}
\put(35.00,5.00){\circle*{5.20}}
\put(45.00,15.00){\circle*{5.20}}
\put(60.00,30.00){\circle*{5.20}}
\put(70.00,40.00){\line(-1,-1){14.00}}
\put(70.00,40.00){\line(1,-1){14.00}}
\put(91.00,19.00){\line(1,-1){14.00}}
\put(35.00,5.00){\line(1,1){14.00}}
\put(70.00,44.00){\makebox(0,0)[cb]{$2\l_n$}}
\put(85.00,30.00){\makebox(0,0)[lc]{$\l_{2\left[\frac{n}{2}\right]-1}$}}
\put(100.00,15.00){\makebox(0,0)[lc]{$\l_3$}}
\put(110.00,5.00){\makebox(0,0)[lc]{$\l_1$}}
\put(30.00,5.00){\makebox(0,0)[rc]{$0$}}
\put(40.00,15.00){\makebox(0,0)[rc]{$\l_2$}}
\put(55.00,30.00){\makebox(0,0)[rc]{$\l_{2\left[\frac{n-1}{2}\right]}$}}
\put(86.00,24.00){\makebox(0,0)[cc]{$\cdot$}}
\put(88.00,22.00){\makebox(0,0)[cc]{$\cdot$}}
\put(90.00,20.00){\makebox(0,0)[cc]{$\cdot$}}
\put(50.00,20.00){\makebox(0,0)[cc]{$\cdot$}}
\put(52.00,22.00){\makebox(0,0)[cc]{$\cdot$}}
\put(54.00,24.00){\makebox(0,0)[cc]{$\cdot$}}
\end{picture}
\caption{The tensor product graph for the product
$V(\l_n)\otimes V(\l_n)$ of two spinor
representations of $B_n$. $\left[\frac{n}{2}\right]$ means the integer
part of $\frac{n}{2}$.\label{bspin}}
\end{figure}
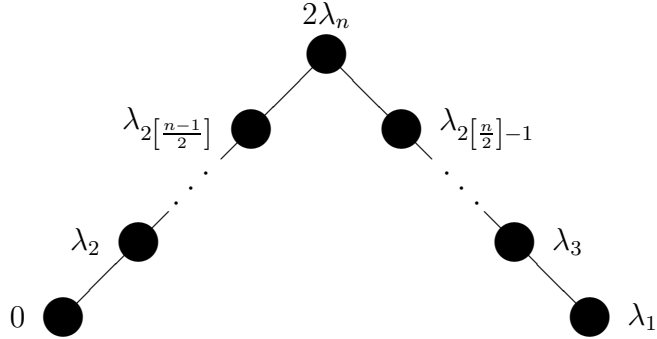
On the representations appearing in this graph the Casimir takes the
values
\beq
C(2\l_n)=n(n+1)~,~~~~
C(\l_a)=a(2n+1-a)~,~~~a=1,\dots,n-1.
\eeq
Using the graph in figure \ref{bspin} and these values of the Casimir,
we can immediately write down the
solution of \reff{rel}. To be explicit we distinguish between the cases
of $n$ even and $n$ odd. For $n$ even we find
\bea
\rho_{\l_{2a}}(x)&=&
\prod_{i=1}^{\frac{n}{2}-a}\left\langle 4i-1\right\rangle\,,
{}~~~~a=0\dots,\frac{n}{2}-1\,,\\
\rho_{\l_{2a+1}}(x)&=&
\prod_{i=1}^{\frac{n}{2}-a}\left\langle 4i-3\right\rangle\,,
{}~~~~a=0\dots,\frac{n}{2}-1\,,
\eea
and for $n$ odd
\bea
\rho_{\l_{2a}}(x)&=&
\prod_{i=1}^{\frac{n+1}{2}-a}\left\langle 4i-3\right\rangle\,,
{}~~~~a=0\dots,\frac{n-1}{2}\,,\\
\rho_{\l_{2a+1}}(x)&=&
\prod_{i=1}^{\frac{n-1}{2}-a}\left\langle 4i-1\right\rangle\,,
{}~~~~a=0\dots,\frac{n-1}{2}-1\,.
\eea
The R-matrix is
\beq
{\bf\check{R}}^{\l_n\l_n}(x)=
\sum_{a=0}^{n-1}\rho_{\l_a}(x){\bf\check{P}}^{\l_n\l_n}_{\l_a}
+ {\bf\check{P}}^{\l_n\l_n}_{2\l_n}.
\eeq

We also read off from the tensor product graph in figure \ref{bspin} that
the module $V(\l_1)$ (i.e., the vector representation)
is affinizable, because it sits on an extremal
node of the graph. Therefore we  can next look at the tensor product
graph for the product of two vector representations
\beq\label{b11}
\unitlength=1mm
\linethickness{0.4pt}
\begin{picture}(102.60,12.60)(10,13)
\put(60.00,20.00){\circle*{5.20}}
\put(80.00,20.00){\circle*{5.20}}
\put(52.00,20.00){\makebox(0,0)[rc]{$V(\l_1)\otimes V(\l_1)~=~$}}
\put(60.00,16.00){\makebox(0,0)[ct]{$0$}}
\put(80.00,16.00){\makebox(0,0)[ct]{$\l_2$}}
\put(100.00,20.00){\circle*{5.20}}
\put(100.00,16.00){\makebox(0,0)[ct]{$2\l_1$}}
\put(60.00,20.00){\line(1,0){40.00}}
\end{picture}
\eeq
{}From this we see that also the module $V(2\l_1)$ (i.e., the symmetric
traceless tensor of rank 2) is affinizable and we can look at the tensor
product involving it. Continuing in this way we see that all the modules
$V(a\l_1)$ for any $a$ are affinizable. Figure \ref{bsym} shows
the tensor product graph for the general product $V(a\l_1)
\otimes V(b\l_1)$ $(a\geq b)$. The modules $V((a+b)\l_1)$ and $V((a-b)\l_1)$
sit on extremal nodes of the graph for $V(a\l_1)\otimes V(b\l_1)$
and are thus affinizable.
\begin{figure}[ht]
\unitlength=1.00mm
\linethickness{0.4pt}
\begin{picture}(129.60,92.00)
\put(47.00,86.00){\line(1,-1){35.00}}
\put(77.00,56.00){\line(-1,0){30.00}}
\put(67.00,51.00){\line(-1,1){20.00}}
\put(47.00,71.00){\line(1,0){15.00}}
\put(47.00,56.00){\line(1,-1){5.00}}
\put(47.00,6.00){\line(1,0){35.00}}
\put(47.00,21.00){\line(1,0){35.00}}
\put(47.00,21.00){\line(1,-1){15.00}}
\put(77.00,6.00){\line(-1,1){20.00}}
\put(127.00,6.00){\line(-1,0){20.00}}
\put(112.00,6.00){\line(-1,1){5.00}}
\put(127.00,6.00){\line(-1,1){20.00}}
\put(112.00,21.00){\line(-1,0){5.00}}
\put(72.00,26.00){\line(1,-1){10.00}}
\put(47.00,86.00){\circle*{5.20}}
\put(47.00,71.00){\circle*{5.20}}
\put(62.00,71.00){\circle*{5.20}}
\put(62.00,56.00){\circle*{5.20}}
\put(47.00,56.00){\circle*{5.20}}
\put(77.00,56.00){\circle*{5.20}}
\put(77.00,21.00){\circle*{5.20}}
\put(62.00,21.00){\circle*{5.20}}
\put(47.00,21.00){\circle*{5.20}}
\put(47.00,6.00){\circle*{5.20}}
\put(62.00,6.00){\circle*{5.20}}
\put(77.00,6.00){\circle*{5.20}}
\put(112.00,6.00){\circle*{5.20}}
\put(127.00,6.00){\circle*{5.20}}
\put(112.00,21.00){\circle*{5.20}}
\put(94.00,21.00){\makebox(0,0)[cc]{$\cdots$}}
\put(94.00,6.00){\makebox(0,0)[cc]{$\cdots$}}
\put(38.00,86.00){\makebox(0,0)[rc]{$(a+b)\l_1$}}
\put(38.00,71.00){\makebox(0,0)[rc]{$(a+b-2)\l_1$}}
\put(38.00,56.00){\makebox(0,0)[rc]{$(a+b-4)\l_1$}}
\put(38.00,21.00){\makebox(0,0)[rc]{$(a-b+2)\l_1$}}
\put(38.00,6.00){\makebox(0,0)[rc]{$(a-b)\l_1$}}
\put(62.00,80.00){\makebox(0,0)[cb]{$\l_2$}}
\put(77.00,65.00){\makebox(0,0)[cb]{$2\l_2$}}
\put(112.00,29.00){\makebox(0,0)[cb]{$(b-1)\l_2$}}
\put(127.00,15.00){\makebox(0,0)[cb]{$b\l_2$}}
\put(38.00,40.00){\makebox(0,0)[rc]{$\vdots$}}
\put(94.00,39.00){\makebox(0,0)[cc]{$\ddots$}}
\put(47.00,92.00){\makebox(0,0)[cb]{$0$}}
\end{picture}
\caption{The extended tensor product graph for the product $V(a\l_1)
\otimes V(b\l_1)$ $(a\geq b)$ of symmetric traceless tensors
of rank $a$ and rank $b$.
The nodes correspond to modules whose highest weight is the sum of the
weight labeling the column and the weight labeling the row.
\label{bsym}}
\end{figure}
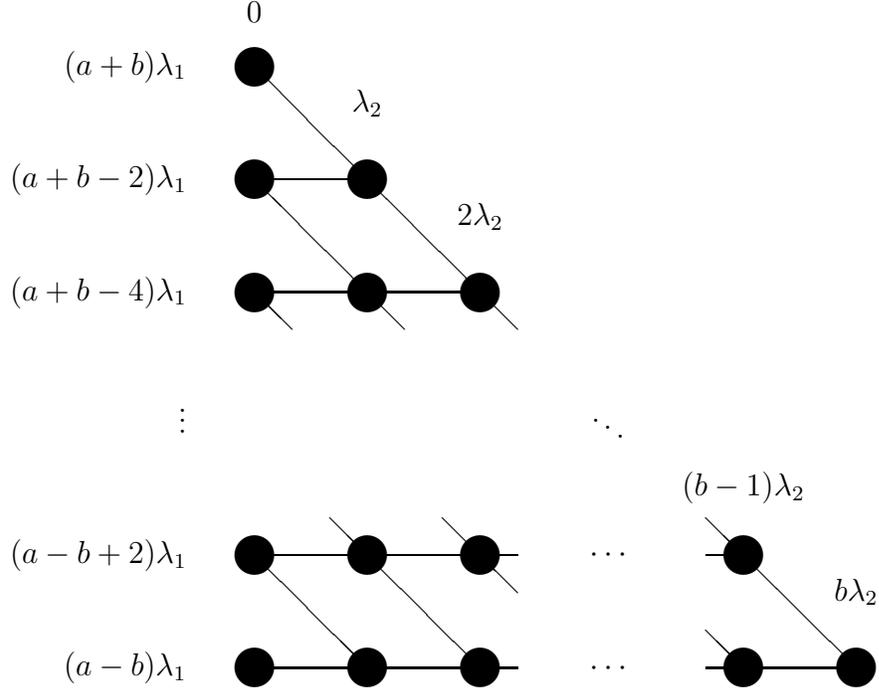

On the modules appearing in figure \ref{bsym} the Casimir takes the values
\beq\label{bsymcas}
C(c\l_1+d\l_2)=c^2+2d^2+2c d+(2n-1)c+(4n-4)d.
\eeq
We need the differences of the values of the Casimir along the edges.
Going down diagonally we get
\bea\label{bsymdiag}
&&C\left((a+b+2-2c)\l_1+(c-d-1)\l_2\right)\nonumber\\
&&-C\left((a+b-2c)\l_1+(c-d)\l_2\right)=2(2+a+b-2c).
\eea
Going from right to left horizontally we get
\bea\label{bsymhori}
&&C\left(a+b-2c)\l_1+(c-d+1)\l_2\right)\nonumber\\
&&-C\left((a+b-2c)\l_1+(c-d)\l_2\right)=2(2n-1+a+b-2d).
\eea

The extended tensor product graph in figure \ref{bsym} is multiply connected.
This means that using relation \reff{rel} one might obtain two different
expressions for the coefficients $\rho_\nu(x)$
if one follows two different paths in
the tensor product graph. One finds however that the values (\ref{bsymcas})
of the Casimir are such that all possible paths give the same result.
This can be seen by observing that the right hand side of \reff{bsymdiag}
is independent of $d$ and the right hand side of \reff{bsymhori} is
independent of $c$. If this had not been the case, then it
would have been necessary to determine which edges of the extended
tensor product graph belong also to the tensor product graph and to use
only those.

{}From figure \ref{bsym} and the Casimir values we read off that
\bea
\rho_{(a+b-2c)\l_1+(c-d)\l_2}(x)
&=& \prod_{i=1}^c\left\langle 2+a+b-2i\right\rangle\\
&&~~\times~\prod_{j=1}^d\left\langle 2n-1+a+b-2j\right\rangle.
\eea
Thus we obtain the R-matrices
\bea
{\bf\check{R}}^{a\l_1,b\l_1}(x)&=&
\sum_{c=0}^b\,\sum_{d=0}^c\,
\prod_{i=1}^c\left\langle 2+a+b-2i\right\rangle~
\prod_{j=1}^d\left\langle 2n-1+a+b-2j\right\rangle\nonumber\\
&&~~~~~~~~~\times{\bf\check{P}}^{a\l_1,b\l_1}_{(a+b-2c)\l_1+(c-d)\l_2}.
\eea

\subsection{Examples for $C_n=sp(2n)$}

The fundamental weights of $C_n$ are
\beq
\l_a=\sum_{i=1}^a\,\e_i\,,~~~~~a=1,\dots,n.
\eeq
and the semisum of positive roots is
\beq
\rho=\sum_{i=1}^n\left(n+1-i\right)\,\e_i.
\eeq

The vector representation $V(\l_1)$ is miniscule and thus affinizable.
Figure \ref{cfund} shows the extended tensor product graph
for the product $V(\l_a)\otimes V(\l_b)$  of two
arbitrary fundamental representations. This graph has been used in
\cite{Mac92a} to determine the rational R-matrices.
By looking at figure \ref{cfund} we can see that all fundamental
representations
are affinizable. This is so because $V(\l_{a+1})$ appears as an extremal
node in $V(\l_a)\otimes V(\l_1)$ for any $a$. We will determine the
trigonometric R-matrix acting in the product $V(\l_a)\otimes V(\l_b)$.

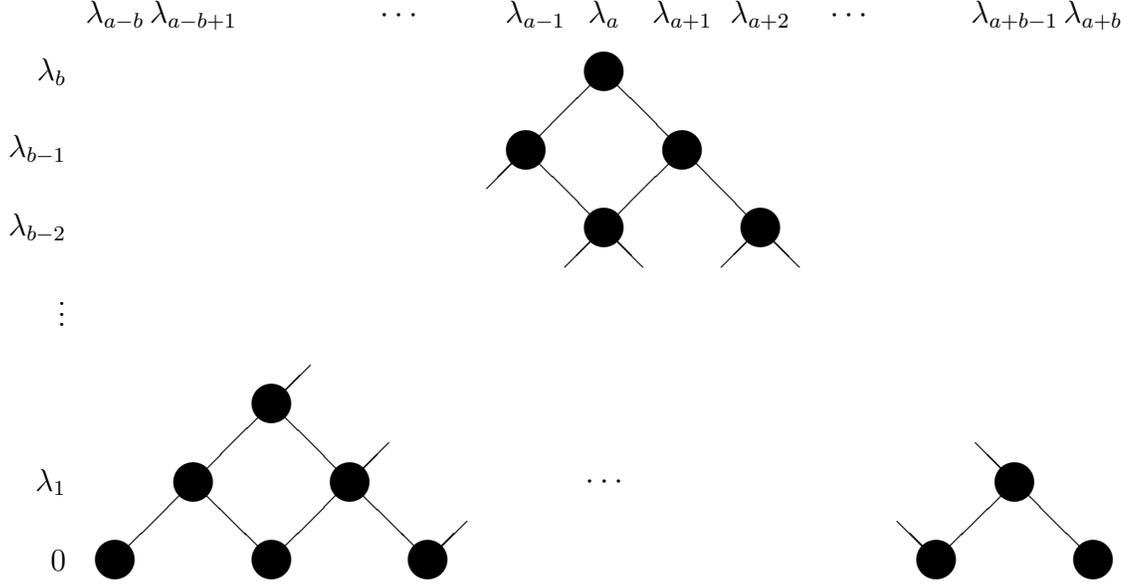
\begin{figure}[ht]
\unitlength=1.3mm
\linethickness{0.4pt}
\begin{picture}(132.00,70.00)(17,3)
\put(30.00,10.00){\circle*{4.00}}
\put(46.00,10.00){\circle*{4.00}}
\put(38.00,18.00){\circle*{4.00}}
\put(54.00,18.00){\circle*{4.00}}
\put(130.00,10.00){\circle*{4.00}}
\put(114.00,10.00){\circle*{4.00}}
\put(122.00,18.00){\circle*{4.00}}
\put(62.00,10.00){\circle*{4.00}}
\put(46.00,26.00){\circle*{4.00}}
\put(80.00,60.00){\circle*{4.00}}
\put(72.00,52.00){\circle*{4.00}}
\put(88.00,52.00){\circle*{4.00}}
\put(96.00,44.00){\circle*{4.00}}
\put(80.00,44.00){\circle*{4.00}}
\put(30.00,10.00){\line(1,1){20.00}}
\put(46.00,10.00){\line(1,1){12.00}}
\put(46.00,26.00){\line(1,-1){16.00}}
\put(38.00,18.00){\line(1,-1){8.00}}
\put(80.00,60.00){\line(-1,-1){12.00}}
\put(80.00,60.00){\line(1,-1){20.00}}
\put(72.00,52.00){\line(1,-1){12.00}}
\put(88.00,52.00){\line(-1,-1){12.00}}
\put(130.00,10.00){\line(-1,1){12.00}}
\put(122.00,18.00){\line(-1,-1){8.00}}
\put(114.00,10.00){\line(-1,1){4.00}}
\put(96.00,44.00){\line(-1,-1){4.00}}
\put(62.00,10.00){\line(1,1){4.00}}
\put(80.00,65.00){\makebox(0,0)[cb]{$\l_a$}}
\put(88.00,65.00){\makebox(0,0)[cb]{$\l_{a+1}$}}
\put(96.00,65.00){\makebox(0,0)[cb]{$\l_{a+2}$}}
\put(105.00,65.00){\makebox(0,0)[cb]{$\cdots$}}
\put(122.00,65.00){\makebox(0,0)[cb]{$\l_{a+b-1}$}}
\put(130.00,65.00){\makebox(0,0)[cb]{$\l_{a+b}$}}
\put(73.00,65.00){\makebox(0,0)[cb]{$\l_{a-1}$}}
\put(30.00,65.00){\makebox(0,0)[cb]{$\l_{a-b}$}}
\put(38.00,65.00){\makebox(0,0)[cb]{$\l_{a-b+1}$}}
\put(59.00,65.00){\makebox(0,0)[cb]{$\cdots$}}
\put(25.00,60.00){\makebox(0,0)[rc]{$\l_b$}}
\put(25.00,52.00){\makebox(0,0)[rc]{$\l_{b-1}$}}
\put(25.00,44.00){\makebox(0,0)[rc]{$\l_{b-2}$}}
\put(25.00,36.00){\makebox(0,0)[rc]{$\vdots$}}
\put(25.00,10.00){\makebox(0,0)[rc]{$0$}}
\put(25.00,18.00){\makebox(0,0)[rc]{$\l_1$}}
\put(80.00,18.00){\makebox(0,0)[cc]{$\cdots$}}
\end{picture}
\caption{The extended tensor product graph for the product $V(\l_a)\otimes
V(\l_b)$ $(a\geq b)$ of two arbitrary fundamental representations of $C_n$.
The nodes correspond to representations whose highest weight is given by
the sum of the weight labeling the column and the weight labeling
the row. If $a+b>n$ then the graph extends to the right only up to
$\l_{n}$.\label{cfund}}
\end{figure}

On the modules appearing in figure \ref{cfund} the Casimir takes the values
\beq\label{cfundcas}
C(\l_a+\l_b)=(2n+2-a)a+(2n+4-b)b.
\eeq
We calculate the differences of the values of the Casimir along the
edges. Going down to the right we get
\beq\label{cfundright}
C(\l_{a+c-d-1}+\l_{b-c-d+1})-C(\l_{a+c-d}+\l_{b-c-d})
=2(a-b+2c).
\eeq
Going down to the left we get
\beq\label{cfundleft}
C(\l_{a+c-d+1}+\l_{b-c-d+1})-C(\l_{a+c-d}+\l_{b-c-d})
=2(2n+2-a-b+2d).
\eeq
Note that again the right hand side of \reff{cfundright} is independent
of $d$ and the right hand side of \reff{cfundleft} is independent of $c$.
This insures that it is irrelevant, which path through the graph we choose
for determining the coefficients $\rho$ using \reff{rel}. We obtain
the following R-matrices
\bea
{\bf\check{R}}^{\l_a,\l_b}(x)&=&
\sum_{c=0}^{min(b,n-a)}\,\sum_{d=0}^{b-c}\,
\prod_{i=1}^c\left\langle a-b+2i\right\rangle~
\prod_{j=1}^d\left\langle 2n+2-a-b+2j\right\rangle\nonumber\\
&&~~~~~~~~~\times{\bf\check{P}}^{\l_a,\l_b}_{\l_{a+c-d}+\l_{b-c-d}}.
\eea
These R-matrices, which are important for the construction of the
S-matrices for solitons in $D_{n+1}^{(2)}$ Toda theory \cite{Ber91},
were first proposed Hollowood \cite{Hol93}. He did however not give a
detailed justification and explanation of his formula.
It was one of the motivations
for the work presented in this paper to derive Hollowood's formula.

\subsection{Examples for $D_n=so(2n)$}

The fundamental weights of $D_n$ are
\beq
\l_n=\half\sum_{i=1}^n \e_i~,~~~
\l_{n-1}=\l_n-\e_n~,~~~~
\l_a=\sum_{i=1}^a \e_i~,~~a=1,\dots,n-2.
\eeq
The semisum of positive roots is
\beq
\rho=\sum_{i=1}^n\,(n-i)\,\e_i.
\eeq

The spinor representations $V(\l_n)$ and $V(\l_{n-1})$ are miniscule
(as well as the vector representation $V(\l_1)$). The tensor product
graphs for the products of two spinor representations are
\beq\label{dspinspin}
\unitlength=1mm
\linethickness{0.4pt}
\begin{picture}(116.60,20.60)(12,2)
\put(114.00,14.00){\circle*{5.20}}
\put(69.00,14.00){\circle*{5.20}}
\put(54.00,14.00){\circle*{5.20}}
\put(48.00,14.00){\makebox(0,0)[rc]{$V(\l_n)\otimes V(\l_n)~=~~$}}
\put(54.00,9.00){\makebox(0,0)[ct]{$2\l_n$}}
\put(69.00,9.00){\makebox(0,0)[ct]{$\l_{n-2}$}}
\put(84.00,14.00){\circle*{5.20}}
\put(84.00,9.00){\makebox(0,0)[ct]{$\l_{n-4}$}}
\put(110.00,9.00){\makebox(0,0)[lt]
{\parbox[t]{3cm}{$0$~ for $n$ even\newline $\l_1$ for $n$ odd}}}
\put(54.00,14.00){\line(1,0){36.00}}
\put(93.00,14.00){\line(1,0){2.00}}
\put(98.00,14.00){\line(1,0){2.00}}
\put(103.00,14.00){\line(1,0){2.00}}
\put(109.00,14.00){\line(1,0){5.00}}
\end{picture}
\eeq
\beq\label{dspinaspin}
\unitlength=1mm
\linethickness{0.4pt}
\begin{picture}(116.60,22.60)(12,2)
\put(114.00,14.00){\circle*{5.20}}
\put(69.00,14.00){\circle*{5.20}}
\put(54.00,14.00){\circle*{5.20}}
\put(48.00,14.00){\makebox(0,0)[rc]{$V(\l_n)\otimes V(\l_{n-1})~=~~$}}
\put(54.00,9.00){\makebox(0,0)[ct]{$\l_n+\l_{n-1}$}}
\put(69.00,9.00){\makebox(0,0)[ct]{$\l_{n-3}$}}
\put(84.00,14.00){\circle*{5.20}}
\put(84.00,9.00){\makebox(0,0)[ct]{$\l_{n-5}$}}
\put(110.00,9.00){\makebox(0,0)[lt]
{\parbox[t]{3cm}{$0$~ for $n$ odd\newline
$\l_1$ for $n$ even}}}
\put(54.00,14.00){\line(1,0){36.00}}
\put(93.00,14.00){\line(1,0){2.00}}
\put(98.00,14.00){\line(1,0){2.00}}
\put(103.00,14.00){\line(1,0){2.00}}
\put(109.00,14.00){\line(1,0){5.00}}
\end{picture}
\eeq
The graph for the product $V(\l_{n-1})\otimes V(\l_{n-1})$ is obtained
from (\ref{dspinspin}) by conjugation (i.e., $\l_n \leftrightarrow \l_{n-1}$).
The Casimir values are
\bea
&&C(\l_a)=a(2n-a)~,~~~~a=1,\dots,n-2\nonumber\\
&&C(\l_n)=n^2~,~~~C(\l_{n-1})=n^2-1.
\eea
This leads to the following formulas for the R-matrices
\bea\label{drspin}
{\bf\check{R}}^{\l_n\l_n}(x)&=&
{\bf\check{P}}^{\l_n\l_n}_{2\l_n}+
\sum_{a=1}^{\left[\frac{n}{2}\right]}\,\prod_{i=1}^a\,
\left\langle 4i-2\right\rangle
{\bf\check{P}}^{\l_n\l_n}_{\l_{n-2a}},\\
{\bf\check{R}}^{\l_n\l_{n-1}}(x)&=&
{\bf\check{P}}^{\l_n\l_{n-1}}_{\l_n+\l_{n-1}}+
\sum_{a=1}^{\left[\frac{n}{2}\right]}\,\prod_{i=1}^a\,
\left\langle 4i\right\rangle
{\bf\check{P}}^{\l_n\l_{n-1}}_{\l_{n-2a}} .
\eea
The R-matrix in \reff{drspin} had already been determined in \cite{Zha91}.

The extended tensor product graph for the product of two symmetric traceless
tensors of arbitrary rank is identical to the corresponding graph
for $B_n$ shown in figure \ref{bsym}. Once again we can show that all
the symmetric traceless tensor representations are affinizable, by
the same argument as before. So we will determine the corresponding
R-matrices. The Casimir takes the values
\beq\label{dsymcas}
C(c\l_1+d\l_2)=c^2+2d^2+2c d+(2n-2)c+(4n-6)d.
\eeq
The differences between the values of the Casimir on connected nodes
are, going down along the diagonal
\bea\label{dsymdiag}
&&C\left((a+b+2-2c)\l_1+(c-d-1)\l_2\right)\nonumber\\
&&-C\left((a+b-2c)\l_1+(c-d)\l_2\right)=2(2+a+b-2c),
\eea
and going left along the horizontal
\bea\label{dsymhori}
&&C\left(a+b-2c)\l_1+(c-d+1)\l_2\right)\nonumber\\
&&-C\left((a+b-2c)\l_1+(c-d)\l_2\right)=2(2n-2+a+b-2d).
\eea
Again the extended tensor product graph leads to a consistent solution of
the relations \reff{rel} and we obtain the R-matrices
\bea
{\bf\check{R}}^{a\l_1,b\l_1}(x)&=&
\sum_{c=0}^b\,\sum_{d=0}^c\,
\prod_{i=1}^c\left\langle 2+a+b-2i\right\rangle~
\prod_{j=1}^d\left\langle 2n-1+a+b-2j\right\rangle\nonumber\\
&&~~~~~~~~~\times{\bf\check{P}}^{a\l_1,b\l_1}_{(a+b-2c)\l_1+(c-d)\l_2}.
\eea

\subsection{Examples for $E_6$}

The Dynkin diagram for $E_6$ is
\beq
\unitlength=1mm
\linethickness{0.4pt}
\begin{picture}(112.00,24.00)(25,3)
\put(50.00,10.00){\circle{4.00}}
\put(65.00,10.00){\circle{4.00}}
\put(80.00,10.00){\circle{4.00}}
\put(95.00,10.00){\circle{4.00}}
\put(110.00,10.00){\circle{4.00}}
\put(80.00,22.00){\circle{4.00}}
\put(52.00,10.00){\line(1,0){11.00}}
\put(67.00,10.00){\line(1,0){11.00}}
\put(82.00,10.00){\line(1,0){11.00}}
\put(97.00,10.00){\line(1,0){11.00}}
\put(80.00,12.00){\line(0,1){8.00}}
\put(50.00,6.00){\makebox(0,0)[ct]{$\alpha_1$}}
\put(65.00,6.00){\makebox(0,0)[ct]{$\alpha_2$}}
\put(80.00,6.00){\makebox(0,0)[ct]{$\alpha_3$}}
\put(95.00,6.00){\makebox(0,0)[ct]{$\alpha_4$}}
\put(110.00,6.00){\makebox(0,0)[ct]{$\alpha_5$}}
\put(84.00,22.00){\makebox(0,0)[lc]{$\alpha_6$}}
\end{picture}
\eeq
We choose
the normalization $(\a_i,\a_i)=2,~i=1,2,\cdots, 6$.
The fundamental weights are
\bea
&&\l_1=\frac{1}{3}(4\a_1+5\a_2+6\a_3+4\a_4+2\a_5+3\a_6),\nonumber\\
&&\l_2=\frac{1}{3}(5\a_1+10\a_2+12\a_3+8\a_4+4\a_5+6\a_6),\nonumber\\
&&\l_3=2\a_1+4\a_2+6\a_3+4\a_4+2\a_5+3\a_6),\nonumber\\
&&\l_4=\frac{1}{3}(4\a_1+8\a_2+12\a_3+10\a_4+5\a_5+6\a_6),\nonumber\\
&&\l_5=\frac{1}{3}(2\a_1+4\a_2+6\a_3+5\a_4+4\a_5+3\a_6),\nonumber\\
&&\l_6=\a_1+2\a_2+3\a_3+2\a_4+\a_5+2\a_6.
\eea
The half sum of positive roots is
\beq
\rho=\sum_{i=1}^6\l_i=8\a_1+15\a_2+21\a_3+15\a_4+8\a_5+11\a_6.
\eeq
The miniscule representations are $V(\l_1)$ and $V(\l_5)$.

We consider the following tensor product graphs
\beq\label{e1}
\unitlength=1.00mm
\linethickness{0.4pt}
\begin{picture}(132.60,12.60)(10,13)
\put(60.00,20.00){\circle*{5.20}}
\put(95.00,20.00){\circle*{5.20}}
\put(52.00,20.00){\makebox(0,0)[rc]{$V(\l_1)\otimes V(a\l_1)~=~$}}
\put(60.00,16.00){\makebox(0,0)[ct]{$(a-1)\l_1+\l_5$}}
\put(95.00,16.00){\makebox(0,0)[ct]{$(a-1)\l_1+\l_2$}}
\put(130.00,20.00){\circle*{5.20}}
\put(130.00,16.00){\makebox(0,0)[ct]{$(a+1)\l_1$}}
\put(60.00,20.00){\line(1,0){70.00}}
\end{picture}
\eeq
We see that the modules $V(a\l_1)$ are affinizable for any $a$ because
$V(a\l_1)$ appears as extremal node on the graph for
$V(\l_1)\otimes V((a-1)\l_1)$.

On the modules appearing in (\ref{e1}) the Casimir takes the values
\bea
&&C((a-1)\l_1+\l_5)=\frac{4}{3}(a-1)^2+\frac{52}{3}(a-1)+
  \frac{52}{3},\nonumber\\
&&C((a-1)\l_1+\l_2)=\frac{4}{3}(a-1)^2+\frac{58}{3}(a-1)+
  \frac{100}{3},\nonumber\\
&&C((a+1)\l_1)=\frac{4}{3}(a+1)^2+16(a+1).
\eea
We obtain the R-matrices
\beq
{\bf\check{R}}^{\l_1,a\l_1}(x)={\bf\check{P}}^{\l_1,a\l_1}_{(a+1)\l_1}
  +\left\langle a+1\right\rangle\,{\bf\check{P}}^{\l_1,a\l_1}_{(a-1)\l_1+\l_2}
  +\left\langle a+1\right\rangle \left\langle a+7\right\rangle\,
  {\bf\check{P}}^{\l_1,a\l_1}_{(a-1)\l_1+\l_5}.
\eeq
The case of $a=1$ was also worked in \cite{Ma90,Zha91}.

Next we look at the tensor product graphs
\beq\label{e2}
\unitlength=1.00mm
\linethickness{0.4pt}
\begin{picture}(132.60,12.60)(10,13)
\put(60.00,20.00){\circle*{5.20}}
\put(95.00,20.00){\circle*{5.20}}
\put(52.00,20.00){\makebox(0,0)[rc]{$V(\l_5)\otimes V(a\l_1)~=~$}}
\put(60.00,16.00){\makebox(0,0)[ct]{$(a-1)\l_1$}}
\put(95.00,16.00){\makebox(0,0)[ct]{$(a-1)\l_1+\l_6$}}
\put(130.00,20.00){\circle*{5.20}}
\put(130.00,16.00){\makebox(0,0)[ct]{$a\l_1+\l_5$}}
\put(60.00,20.00){\line(1,0){70.00}}
\end{picture}
\eeq
The Casimir values are
\bea
&&C((a-1)\l_1)=\frac{4}{3}(a-1)^2+16(a-1),\nonumber\\
&&C(a\l_1+\l_5)=\frac{4}{3}a^2+\frac{52}{3}a+\frac{52}{3},\nonumber\\
&&C((a-1)\l_1+\l_6)=\frac{4}{3}(a-1)^2+18(a-1)+24.
\eea
We obtain the R-matrices
\beq
{\bf\check{R}}^{\l_5,a\l_1}(x)={\bf\check{P}}^{\l_5,a\l_1}_{a\l_1+\l_5}
  +\left\langle 2(a+1)\right\rangle \,{\bf\check{P}}^{\l_5,a\l_1}_{(a-1)
  \l_1+\l_6}+\left\langle 2(a+1)\right\rangle\left\langle a+11\right\rangle\,
  {\bf\check{P}}^{\l_5,a\l_1}_{(a-1)\l_1}.
\eeq

\subsection{Example for $F_4$}

The Dynkin diagram for $F_4$ is
\beq
\unitlength=1mm
\linethickness{0.4pt}
\begin{picture}(97.00,8.00)(30,10)
\put(50.00,15.00){\circle{4.00}}
\put(65.00,15.00){\circle{4.00}}
\put(80.00,15.00){\circle{4.00}}
\put(95.00,15.00){\circle{4.00}}
\put(93.00,15.00){\line(-1,0){11.00}}
\put(78.00,16.00){\line(-1,0){11.00}}
\put(67.00,14.00){\line(1,0){11.00}}
\put(63.00,15.00){\line(-1,0){11.00}}
\put(50.00,12.00){\makebox(0,0)[ct]{$\alpha_1$}}
\put(65.00,12.00){\makebox(0,0)[ct]{$\alpha_2$}}
\put(80.00,12.00){\makebox(0,0)[ct]{$\alpha_3$}}
\put(95.00,12.00){\makebox(0,0)[ct]{$\alpha_4$}}
\put(95.00,18.00){\makebox(0,0)[cb]{$1$}}
\put(80.00,18.00){\makebox(0,0)[cb]{$1$}}
\put(65.00,18.00){\makebox(0,0)[cb]{$2$}}
\put(50.00,18.00){\makebox(0,0)[cb]{$2$}}
\end{picture}
\eeq
We choose the normalization
$(\a_1,\a_1)=(\a_2,\a_2)=2(\a_3,\a_3)=2(\a_4,\a_4)=4$ so that
$(\a_1,\a_2)=(\a_2,\a_1)=(\a_2,\a_3)=(\a_3,\a_2)=2(\a_3,\a_4)=2(\a_4,\a_3)=-2$
with $(\a_i,\a_j)=0$ for all other pairs. The fundamental weights are
\bea
&&\l_1=2\a_1+3\a_2+4\a_3+2\a_4\,,\nonumber\\
&&\l_2=3\a_1+6\a_2+8\a_3+4\a_4\,,\nonumber\\
&&\l_3=2\a_1+4\a_2+6\a_3+3\a_4\,,\nonumber\\
&&\l_4=\a_1+2\a_2+3\a_3+2\a_4\,.
\eea
The half sum of positive roots is
\beq
\rho=\sum_{i=1}^4\l_i=8\a_1+15\a_2+21\a_3+11\a_4.
\eeq

It can be shown that the minimal representation $V(\l_4)$ is affinizable.
Thus we consider the following tensor product graph
\beq
\unitlength=1mm
\linethickness{0.4pt}
\begin{picture}(112.60,9.60)(10,10)
\put(50.00,15.00){\circle*{5.20}}
\put(65.00,15.00){\circle*{5.20}}
\put(80.00,15.00){\circle*{5.20}}
\put(95.00,15.00){\circle*{5.20}}
\put(110.00,15.00){\circle*{5.20}}
\put(110.00,15.00){\line(-1,0){60.00}}
\put(44.00,15.00){\makebox(0,0)[rc]{$V(\l_4)\otimes V(\l_4)~=$}}
\put(50.00,11.00){\makebox(0,0)[ct]{$0$}}
\put(65.00,11.00){\makebox(0,0)[ct]{$\l_1$}}
\put(80.00,11.00){\makebox(0,0)[ct]{$2\l_4$}}
\put(95.00,11.00){\makebox(0,0)[ct]{$\l_3$}}
\put(110.00,11.00){\makebox(0,0)[ct]{$\l_4$}}
\end{picture}
\eeq
On the modules appearing in the above diagram the Casimir has the values
\beq
C(\l_1)=36,~~~C(2\l_4)=52,~~~C(\l_3)=48,~~~C(\l_4)=24
\eeq
from which we obtain the R-matrix
\bea
{\bf\check{R}}^{\l_4\l_4}(x)&=&{\bf\check{P}}^{\l_4\l_4}_0+\left\langle -18
  \right\rangle\,{\bf\check{P}}^{\l_4\l_4}_{\l_1}+\left\langle -18
  \right\rangle\left\langle -8\right\rangle\,
  {\bf\check{P}}^{\l_4\l_4}_{2\l_4}\nonumber\\
& &  +\left\langle -18\right\rangle\left\langle -8\right\rangle\left\langle 2
  \right\rangle\,{\bf\check{P}}^{\l_4\l_4}_{\l_3}+\left\langle -18
  \right\rangle\left\langle -8\right\rangle\left\langle 2\right\rangle
  \left\langle 12\right\rangle\,
  {\bf\check{P}}^{\l_4\l_4}_{\l_4}.
\eea

\sect{Discussion\label{discussion}}

In this paper we have presented a systematic procedure for obtaining
trigonometric solutions $R(x)\in {\rm End} (V(\l)\otimes V(\mu))$ to the
QYBE for a quantum simple Lie algebra $U_q({\cal G})$, where $V(\l),~
V(\mu)$ are irreducible $U_q({\cal G})$-modules which are affinizable and whose
tensor product decomposition is multiplicity-free. Our method relies on
the extended tensor product graph and certain subgraphs. An important
role is played by parities, given by the eigenvalues of the permutation
operator for the case $\l=\mu$, and which we extended to the case $\l\neq\mu$
in the paper. As noted in the introduction spectral dependent R-matrices
obtained this way are central to the theory of quantum integrable systems
occuring in statistical mechanics, the quantum inverse scattering method
and quantum field theory.

Our approach, which is a generalization of that proposed in \cite{Zha91}
for the case $\l=\mu$, enables the construction of a large number of new
R-matrices. As explicit examples we have considered the case where
$V(\l),~V(\mu)$ correspond to the symmetric or antisymmetric tensor
representations for $\G=A_n$, the spinor irreps or the symmetric traceless
tensor irreps for $\G=B_n,~D_n$, all fundamental irreps for $\G=C_n$ and
certain irreps for the exceptional Lie algebras $\G=E_6,~F_4$.

It should be emphasized that although all finite dimensional irreps of
$U_q(\G)$ are known to be affinizable for the case $\G=A_n$, this is not
true in general for other finite dimensional simple Lie algebras.
In this paper we have concentrated on affinizable representations.
There are however many more finite-dimensional irreducible representations
of $U_q(\hat{\G})$ which are obtained not by affinizing $U_q({\G})$ irreducible
representations but by affinizing $U_q(\G)$ reducible ones.
These can be constructed by reduction of tensor product
representations \cite{Del94}. In future work we aim to apply the technique
of this paper to obtain the trigonometric R-matrices in these representations.
They are important for example for the construction of soliton S-matrices
in affine Toda theories.

Finally it is of interest to consider extensions to quantum superalgebras
of relevance to supersymmetric quantum integrable systems. Particularly
interesting are the type-I quantum superalgebras which admit one parameter
families of finite dimensional irreps \cite{Lin92}. The R-matrices arising
from these irreps thus automatically contain an extra non-additive parameter
\cite{Bracken94} corresponding to one parameter families of exactly solvable
models.

\vskip.3in
\begin{center}
{\bf Acknowledgements:}
\end{center}
We thank Anthony J. Bracken and Jon R. Links for discussions. G.W.D. thanks
the Math department of the University of Queensland for hospitality during his
visit. M.D.G. and Y.Z.Z. are financially supported by
Australian Research Council.

\appendix

\sect{Normalization Convention\label{u-condition}}

We have seen that for the universal $R$-matrix $R$ for $U_q(\hat{\G})$,
\beq
R(x)\equiv (D_x\otimes I)(R)
\eeq
so that in general
\beq
R^{\l\mu}(x)=g_q(x) (\pi_\l\otimes\pi_\mu)(R(x))
\eeq
which follows from the uniqueness of solutions to JEs. Here $g_q(x)$ is an
overall scalar function. We thus have in particular
\beq
\lim_{x\rightarrow 0}R^{\l\mu}(x)=g_q(0) (\pi_\l\otimes\pi_\mu)(\bar{R})
\label{a1}
\eeq
for some non-zero $g_q(0)$, with $\bar{R}\equiv R(0)$ the universal
$R$-matrix for $U_q(\G)$.

By Schur's lemma, the $R$-matrix defined above
satisfies the unitarity condition
\beq
(R^T)^{\l\mu}(x)R^{\l\mu}(x^{-1})=c_q(x)\cdot I
\eeq
where $c_q(x)$ is some overall scalar function, due to the
irreducibility of the representation $(\pi_\l\otimes\pi_\mu)
(D_x\otimes 1)$ for generic $x$
and the fact that the left hand side of above equation intertwines the
coproduct of $U_q(\hat{\G})$.

In the context we assume $R^{\l\mu}(x)$ normalized so that
\beq
(R^T)^{\l\mu}(x)R^{\l\mu}(x^{-1})=I.\label{a2}
\eeq
However this does not uniquely specify $R^{\l\mu}(x)$ since it may be
multiplied by an arbitrary function $f_q(x)$ such that
\beq
f_q(x)\cdot f_q(x^{-1})=1.
\eeq
In particular, choosing
\beq
f_q(x)=\frac{x+g_q(0)^{-1}}{1+g_q(0)^{-1}x}
\eeq
we obtain an $R$-matrix  $f_q(x)R^{\l\mu}(x)$ satisfying both equation
(\ref{a2}) as well as
\beq
\lim_{x\rightarrow 0}f_q(x)R^{\l\mu}(x)=(\pi_\l\otimes\pi_\mu)(\bar{R}).
\eeq

Hence without loss of generality we may assume $R^{\l\mu}(x)$
satisfies (\ref{a2}) \underline{and} the limiting condition
\beq
\lim_{x\rightarrow 0}R^{\l\mu}(x)=(\pi_\l\otimes\pi_\mu)(\bar{R})
\eeq
as was assumed above in the context.

It is also important to note that the universal $R$-matrix $\bar{R}$
for $U_q(\G)$ is normalized so that
\beq
\lim_{q\rightarrow 1}\bar{R}=I
\eeq
and
\beq
\D(\bar{R}^T\bar{R})=(v\otimes v)\D(v^{-1})
\eeq
where $v$ is the canonical Casimir with the eigenvalue $q^{-(\l,\l+2\rho)}$ on
the irreducible $U_q(\G)$-module $V(\l)$.

\sect{Parity\label{parity}}

In this appendix we will clarify the parities of module $V(\nu)$ in
$V(\l)\otimes V(\mu)$ by finding a parity operator in this case.

{}From the properties of the "projections" ${\bf\check{P}}^{\l\mu}_\nu$ we
note that
\beq
{\bf\check{R}}^{\l\mu}(1){\bf\check{P}}^{\mu\l}_\nu={\bf\check{P}}^{\l\mu}_\nu
  {\bf\check{R}}^{\mu\l}(1)
\eeq
from which it follows
\beq
{\bf\check{R}}^{\l\mu}(1){\bf\check{R}}^{\mu\l}(x)={\bf\check{R}}^{\l\mu}(x)
{\bf\check{R}}^{\mu\l}(1).
\eeq
We have
\beq
(R^T)^{\l\mu}(1)R^{\l\mu}(x)={\bf\check{R}}^{\mu\l}(1){\bf\check{R}}^{\l\mu}(x)=
 \sum_\nu\rho_\nu(x)\P^{\l\mu}_\nu\label{b1}
 \eeq
 which gives rise to
 \beq
 R^{\l\mu}(x)=R^{\l\mu}(1)\sum_\nu\rho_\nu(x)\P^{\l\mu}_\nu
 \eeq
 which gives a direct expression for $R^{\l\mu}(x)$ on $V(\l)\otimes V(\mu)$.
 It appears therefore that $R^{\l\mu}(1)$ plays a role on
 $V(\l)\otimes V(\mu)$ analogous to the permutation operator when $\l=\mu$.

Taking the limit $q\rightarrow 1$:
\beq
\left .\left .R^{\l\mu}(0)\right |_{q=1}\equiv \bar{R}^{\l\mu}\right |_{q=1}
  =\left .(\pi_\l\otimes \pi_\mu)
  (\bar{R})\right |_{q=1}=I
\eeq
thus  by (\ref{b1})
\beq
\left .(R^T)^{\l\mu}(1)\right |_{q=1}=\sum_\nu\e(\nu)\P^{(0)\l\mu}_\nu
\eeq
where $\P^{(0)\l\mu}_\nu=\P^{\l\mu}_\nu|_{q=1}$.
Since $(R^T)^{\l\mu}(1)=R^{\l\mu}(1)^{-1}$ it follows that at $q=1$
\beq
R^{\l\mu}(1)=\sum_\nu\e(\nu)\P^{(0)\l\mu}_\nu
\eeq
which squares to the identity:
\beq
R^{\l\mu}(1)^2=I,~~~~~~~~~{\rm at}~q=1
\eeq

It thus follows that the parities are determined by the eigenvalues of the
operator
\beq
P\equiv R^{\l\mu}(1)|_{q=1}, ~~~~~P^2=I
\eeq
on the space $V(\l)\otimes V(\mu)$ at $q=1$. It remains to give a natural
geometric interpretation of this parity operator which, in the case $\l=\mu$,
concides with the usual permutation operator. This is the problem addressed
below.

It follows from (\ref{p-0}), (\ref{p-infty}) and (\ref{p-1}) that
for $\nu\neq\nu'$
\beq
{\bf\check{P}}^{\l\mu}_\nu\left (\pi_\l(e_0)\otimes \pi_\mu(\qh)
  \right )\P^{\l\mu}_{\nu'}
  =\P^{\mu\l}_\nu\left (\pi_\mu(e_0)\otimes \pi_\l(\qh)
  \right ){\bf\check{P}}^{\l\mu}_{\nu'}.
  \label{p-1'}
\eeq
which gives rise to
\beq
R^{\l\mu}(1)\P^{\l\mu}_\nu\left (\pi_\l(e_0)\otimes \pi_\mu(\qh)
  \right )\P^{\l\mu}_{\nu'}
  =\bar{\P}^{\mu\l}_\nu\left (\pi_\l(\qh)\otimes\pi_\mu(e_0)
  \right )\bar{\P}^{\mu\l}_{\nu'}R^{\l\mu}(1)
\eeq
where $\bar{\P}^{\l\mu}\equiv P^{\mu\l}\P^{\mu\l}_\nu P^{\l\mu}$ is the
projection onto submodule $\bar{V}(\nu)$ of $V(\l)\otimes V(\mu)$ determined
by the opposite coproduct $\D^{'\l\mu}$ on the tensor product space.

We now consider solely the case $q=1$.
In this case the coproduct and the opposite coproduct coincide so the
above equation reduces to (omitting $\pi_\l$ and $\pi_\mu$ below)
\beq
P\P^{(0)\l\mu}_\nu\left (e_0\otimes 1
  \right )\P^{(0)\l\mu}_{\nu'}
  =\P^{(0)\l\mu}_\nu\left (1\otimes e_0
  \right )\P^{(0)\l\mu}_{\nu'}P\,,~~~~\nu\neq\nu'
\eeq
and, since $P^2=1$, we similarily have
\beq
P\P^{(0)\l\mu}_\nu\left (1\otimes e_0
  \right )\P^{(0)\l\mu}_{\nu'}
  =\P^{(0)\l\mu}_\nu\left (e_0\otimes 1
  \right )\P^{(0)\l\mu}_{\nu'}P\,,~~~~\nu\neq\nu'.
\eeq
Moreover, since $P$ is an $\G$-invariant we must have, for $\nu\neq\nu'$
\bea
&&P\P^{(0)\l\mu}_\nu\left (a\otimes 1
  \right )\P^{(0)\l\mu}_{\nu'}
  =\P^{(0)\l\mu}_\nu\left (1\otimes a
  \right )\P^{(0)\l\mu}_{\nu'}P\nonumber\\
&&P\P^{(0)\l\mu}_\nu\left (1\otimes a
  \right )\P^{(0)\l\mu}_{\nu'}
  =\P^{(0)\l\mu}_\nu\left (a\otimes 1
  \right )\P^{(0)\l\mu}_{\nu'}P,~~~\forall a\in\G.
\eea
We now observe that $a\otimes 1-1\otimes a$ \underline{reverses} the
parity since
\beq
P\P^{(0)\l\mu}_\nu\left (a\otimes 1-1\otimes a
  \right )\P^{(0)\l\mu}_{\nu'}
  =-\P^{(0)\l\mu}_\nu\left (a\otimes 1-1\otimes a
  \right )\P^{(0)\l\mu}_{\nu'}P\,,~~~~\nu\neq\nu'.
\eeq
This suggests we define a sequence of $\G$-modules $V_i,~\bar{V}_i$
recursely according to
\bea
&&\bar{V}_i=\bar{V}_{i-1}\bigoplus V_i,~~(V_i{\rm ~the~orthocomplement~of}~
  \bar{V}_{i-1}~{\rm in}~\bar{V}_i)\nonumber\\
&&\bar{V}_i={\rm span}\{(a\otimes 1-1\otimes a)v|v\in\bar{V}_{i-1},~
  a\in \G\}
\eea
starting with $\bar{V}_0=V_0=V(\l+\mu)$. We then obtain a $\G$-module
direct sum decomposition
\beq
V(\l)\otimes V(\mu)=\bigoplus_{i=0}^k V_i
\eeq
where $\bar{V}_k\equiv V(\l)\otimes V(\mu)$ is uniquely determined by
\beq
(a\otimes 1-1\otimes a)\bar{V}_k\subseteq \bar{V}_k,~~~\forall a\in\G
\eeq

We assume the highest component
$V(\l+\mu)$ with maximal weight vector $v^\l_+\otimes v^\mu_+$ has
positive parity (by convention).
The submodules with positive and negative parities are then given respectively
by
\bea
&&V_+=V_0\bigoplus V_2\bigoplus \cdots\nonumber\\
&&V_-=V_1\bigoplus V_3\bigoplus \cdots
\eea
It follows that
\beq
V(\l)\otimes V(\mu)=V_+\bigoplus V_-.
\eeq
Thus we have given a method how to determine the parity of $V(\nu)\subset
V(\l)\otimes V(\mu)$, which, with the help of $R^{\l\mu}(1)$, is analogous
to the case of $\l=\mu$.

We are now to give a necassary condition for eq.(\ref{me}).
Given two irreducible modules
$V(\nu),~V(\nu')\subseteq V(\l)\otimes V(\mu)$, we have
\bea
\P^{(0)\l\mu}_\nu(e_0\otimes 1)\P^{(0)\l\mu}_{\nu'}&=&\frac{1}{2}
 \P^{(0)\l\mu}_\nu(e_0\otimes 1+1\otimes e_0)\P^{(0)\l\mu}_{\nu'}\nonumber\\
& &+ \frac{1}{2} \P^{(0)\l\mu}_\nu(e_0\otimes 1-1\otimes e_0)
  \P^{(0)\l\mu}_{\nu'}\nonumber\\
&=&\frac{1}{2}  \P^{(0)\l\mu}_\nu(e_0\otimes 1-1\otimes e_0)
  \P^{(0)\l\mu}_{\nu'},~~~~\forall \nu\neq\nu'
\eea
Thus, for $\nu\neq\nu'$, the left hand side of the above equation
can be non-vanishing only if $\nu,~\nu'$ have opposite parity. Therefore
following the similar arguments to \cite{Zha91} we have
\vskip.1in
\noindent{\bf Proposition:} for any $\nu\neq\nu'$, ~
$\P^{\l\mu}_\nu\left (\pi_\l(e_0)\otimes \pi_\mu(\qh)\right )
  \P^{\l\mu}_{\nu'}\neq 0$ only if the parities $\e(\nu),~\e(\nu')$
associated respectively with $\nu,~\nu'$ are opposite, i.e.
$\e(\nu)\,\e(\nu')=-1$.

\sect{``Projectors"\label{projectors}}

In this appendix we show that the ``projectors" defined in the paper
may be determined by using pure representation theory of $U_q(\G)$.
We only list the results below. The details will be published in a
forthcoming paper \cite{DGZ94}.

Let $\{\left |e^\nu_\a(q)\right\rangle_{\l\otimes\mu}\}$ be a symmetry adapted
orthonormal basis for $V(\nu)\subset
V(\l)\otimes V(\mu)$ under the action defined by the coproduct $\D$.
Then obviously $\P^{\l\mu}_{\nu}$ may expressed as
\beq
\P^{\l\mu}_\nu=\sum_\a \left |e^\nu_\a(q)\right\rangle_{\l\otimes\mu~
  \l\otimes\mu}\!\left\langle e^\nu_\a(q)\right |
\eeq
Moreover, it can be shown that
$\{\left |e^\nu_\a(q^{-1})\right\rangle_{\l\otimes\mu}\}$ describes a symmetry
adapted orthonormal basis under the action defined by the opposite
coproduct $\D'$.

We define the symmetry adapted basis for
$V(\nu)\subset V(\mu)\otimes V(\l)$ by
\beq
\left |e^\nu_\a(q)\right\rangle_{\mu\otimes\l}=\e(\nu)P^{\l\mu}
  \left |e^\nu_\a(q^{-1})\right\rangle_{\l\otimes\mu},
\eeq
which is easily seen to be orthonormal so that
\beq
{}_{\mu\otimes \l}\!\left\langle e^\nu_\a(q)\right |
  =\e(\nu)_{~\l\otimes\mu}\!\left\langle e^\nu_\a(q^{-1})\right |P^{\mu\l}\,.
\eeq
It can be proven that for real generic $q>0$, the ``projectors"
${\bf\check{P}}^{\l\mu}_\nu$ may be written as
\beq
{\bf\check{P}}^{\l\mu}_\nu=\sum_\a\left
|e^\nu_\a(q)\right\rangle_{\mu\otimes\l~
  \l\otimes\mu}\!\left\langle e^\nu_\a(q)\right |
\eeq
This implies that for real generic $q>0$ the operators
${\bf\check{P}}^{\l\mu}_\nu$
defined in section \ref{general} can be computed by pure
representation theory (say, CG coefficients) of $U_q(\G)$.

By analytic continuation, the above result should extend to all complex $q$.

\renewcommand{\baselinestretch}{0.9} 
\newcommand{\bib}{\vspace{-2mm}

        \bibitem}

\end{document}